\documentstyle{mn}
\input psfig.sty
\newif\ifAMStwofonts

\newcommand{\figstart}[1]  {\begin{figure} \psfig{#1}}
\newcommand{\lfigstart}[1] {\begin{figure*} \psfig{#1}}
\newcommand{\figend} {\end{figure}}
\newcommand{\lfigend} {\end{figure*}}

\ifoldfss
  \ifCUPmtlplainloaded \else
    \NewTextAlphabet{textbfit} {cmbxti10} {}
    \NewTextAlphabet{textbfss} {cmssbx10} {}
    \NewMathAlphabet{mathbfit} {cmbxti10} {} 
    \NewMathAlphabet{mathbfss} {cmssbx10} {} 
  \fi
  \ifAMStwofonts
    \ifCUPmtlplainloaded \else
      \NewSymbolFont{upmath} {eurm10}
      \NewSymbolFont{AMSa} {msam10}
      \NewMathSymbol{\upi}     {0}{upmath}{19}
      \NewMathSymbol{\umu}     {0}{upmath}{16}
      \NewMathSymbol{\upartial}{0}{upmath}{40}
      \NewMathSymbol{\leqslant}{3}{AMSa}{36}
      \NewMathSymbol{\geqslant}{3}{AMSa}{3E}

       \let\le=\leqslant
       
    \fi
  \fi
\fi 

\ifnfssone
  \newmathalphabet{\mathit}
  \addtoversion{normal}{\mathit}{cmr}{m}{it}
  \addtoversion{bold}{\mathit}{cmr}{bx}{it}
  \newmathalphabet{\mathbfit} 
  \addtoversion{normal}{\mathbfit}{cmr}{bx}{it}
  \addtoversion{bold}{\mathbfit}{cmr}{bx}{it}
  \newmathalphabet{\mathbfss} 
  \addtoversion{normal}{\mathbfss}{cmss}{bx}{n}
  \addtoversion{bold}{\mathbfss}{cmss}{bx}{n}
  \ifAMStwofonts
    \ifCUPmtlplainloaded \else
      \UseAMStwoboldmath
      \makeatletter
      \new@mathgroup\upmath@group
      \define@mathgroup\mv@normal\upmath@group{eur}{m}{n}
      \define@mathgroup\mv@bold\upmath@group{eur}{b}{n}
      \edef\UPM{\hexnumber\upmath@group}
      \new@mathgroup\amsa@group
      \define@mathgroup\mv@normal\amsa@group{msa}{m}{n}
      \define@mathgroup\mv@bold\amsa@group{msa}{m}{n}
      \edef\AMSa{\hexnumber\amsa@group}
      \makeatother
      \mathchardef\upi="0\UPM19
      \mathchardef\umu="0\UPM16
      \mathchardef\upartial="0\UPM40
      \mathchardef\leqslant="3\AMSa36
      \mathchardef\geqslant="3\AMSa3E

       \let\le=\leqslant

    \fi
  \fi
\fi 

\ifnfsstwo
  \DeclareMathAlphabet{\mathbfit}{OT1}{cmr}{bx}{it}
  \SetMathAlphabet\mathbfit{bold}{OT1}{cmr}{bx}{it}
  \DeclareMathAlphabet{\mathbfss}{OT1}{cmss}{bx}{n}
  \SetMathAlphabet\mathbfss{bold}{OT1}{cmss}{bx}{n}
  \ifAMStwofonts
    \ifCUPmtlplainloaded \else
      \DeclareSymbolFont{UPM}{U}{eur}{m}{n}
      \SetSymbolFont{UPM}{bold}{U}{eur}{b}{n}
      \DeclareSymbolFont{AMSa}{U}{msa}{m}{n}
      \DeclareMathSymbol{\upi}{0}{UPM}{"19}
      \DeclareMathSymbol{\umu}{0}{UPM}{"16}
      \DeclareMathSymbol{\upartial}{0}{UPM}{"40}
      \DeclareMathSymbol{\leqslant}{3}{AMSa}{"36}
      \DeclareMathSymbol{\geqslant}{3}{AMSa}{"3E}

       \let\le=\leqslant

    \fi
  \fi
\fi 

\ifCUPmtlplainloaded \else
  \ifAMStwofonts \else 
    \def\upi{\pi}
    \def\umu{\mu}
    \def\upartial{\partial}
  \fi
\fi

\def\hmpc{h^{-1}{\rm Mpc}}
\def\hkpc{$h^{-1}{\rm kpc}$}

\def\msun{M_\odot}

\def\hmsun{h^{-1}M_\odot}
\def\lcdm{$\Lambda$CDM}
\def\tcdm{$\tau$CDM}
\def\scdm{SCDM}
\def\ocdm{OCDM}
\def\etal{{\it et al}}

\title[Mass function of dark matter halos]
{The mass function of dark matter halos} 
\author[A. Jenkins et al.]
{A.~Jenkins,$^1$ C.~S.~Frenk,$^1$ S. D. M.~White,$^2$ J.~M.~Colberg,$^2$\\  
\newauthor S.~Cole,$^1$ A.~E.~Evrard,$^3$ H. M. P. Couchman$^4$ and N.~Yoshida.$^2$ \\
$^1$ Dept Physics, University of Durham, South Road, Durham, DH1 3LE \\
$^2$ Max-Planck Inst. for Astrophysics, Garching, Munich, D-85740, Germany\\
$^3$ Dept Physics, University of Michigan, Ann Arbor, MI-48109-1120.\\
$^4$ Dept Physics \& Astronomy, McMaster University, Hamilton, Ontario, LS8 4M1, Canada}
\date{As accepted by MNRAS}
\pagerange{\pageref{firstpage}--\pageref{lastpage}}
\pubyear{2000}

\begin{document}
\label{firstpage}

\maketitle

\begin{abstract}

We combine data from a number of N-body simulations to 
predict the abundance of dark  halos in Cold Dark Matter 
universes over more than 4 orders of magnitude in mass.
A comparison of different simulations suggests that the dominant 
uncertainty in our results is systematic and is smaller than 
10--30\% at all masses, depending on the halo definition used. 
In particular, our ``Hubble Volume'' simulations of 
\tcdm\ and \lcdm\ cosmologies allow the abundance of massive
clusters to be predicted with uncertainties well below those 
expected in all currently planned observational surveys.
We show that for a range of CDM cosmologies and for
a suitable halo definition, the simulated mass function
is almost independent of epoch, of cosmological parameters, and of initial
power spectrum when expressed in appropriate variables. This
universality is of exactly the kind predicted by the familiar
Press-Schechter model, although this model predicts a mass function
shape which differs from our numerical results, overestimating the 
abundance of ``typical'' halos and underestimating that
of massive systems.

\end{abstract}

\section{Introduction}

Accurate theoretical predictions for halo mass functions are needed 
for a number of reasons. For example, they are a primary input for 
modelling galaxy formation, since current theories assume galaxies 
to result from the condensation of gas in halo cores (White \& Rees 
1978 and many subsequent papers; see Somerville \& Primack 1999 for 
a recent overview). The abundance of the most massive halos is 
sensitive to the overall amplitude of mass fluctuations while the 
evolution of this abundance is sensitive to the cosmological density 
parameter, $\Omega_0$. As a result, identifying massive halos with 
rich galaxy clusters can provide an estimate both of the amplitude 
of the primordial density fluctuations and of the value of $\Omega_0$;
recent discussions include Henry (1997), Eke \etal\ (1998), Bahcall 
\& Fan (1998), Blanchard \& Bartlett (1998) and the Virgo consortium 
presentation of the Hubble Volume simulations used below (Evrard 
\etal\ 2000). The mass function is also a critical ingredient in
the apparently strong constraints on cosmological parameters (principally
$\Omega_0$ and $\Lambda$) which can be derived from the observed
incidence of strong gravitational lensing (e.g. Bartelmann \etal\
1998; Falco, Kochanek \& Munoz 1998). 

As the observational data relevant to these issues improve, the need
for accurate theoretical predictions increases. By far the most widely
used analytic formulae for halo mass functions are based on extensions
of the theoretical framework first sketched by Press \& Schechter 
(1974). Unfortunately, none of these derivations is sufficiently
rigorous that the resulting formulae can be considered accurate
beyond the regime where they have been tested against 
N-body simulations.  In this paper, we combine mass functions from
simulations of four popular versions of the cold dark matter (CDM) 
cosmology to obtain results valid over a wider mass range and to higher 
accuracy than has been possible before.  These mass functions show a 
regularity which allows them all to be fitted by a single interpolation 
formula despite the wide range of epochs and masses they cover.
This formula can be used to obtain accurate predictions for CDM models
with parameters other than those we have simulated explicitly. 

Although the analytical framework of the Press-Schechter (P-S) model 
has been greatly refined and extended in recent years, in particular
to allow predictions for the merger histories of dark
matter halos (Bond \etal\ 1991, Bower 1991, Lacey \& Cole 1993), 
it is well known that the P-S mass function, while qualitatively
correct, disagrees in detail with the results of
N-body simulations. Specifically, the P-S formula overestimates the 
abundance of halos near the characteristic mass $M_*$ and
underestimates the abundance in the high mass tail (Efstathiou \etal\ 
1988, Efstathiou \& Rees 1988, White, Efstathiou \& Frenk 1993,
Lacey \& Cole 1994, Eke, Cole \& Frenk 1996, Gross \etal\ 1998, 
Governato \etal\ 1999).  Recent work has studied whether this
discrepancy can be resolved by replacing the spherical collapse model
of the standard P-S analysis by 
ellipsoidal collapse (e.g. Monaco 1997ab, Lee \& Shandarin 1998,
Sheth, Mo \& Tormen 1999). Sheth, Mo \& Tormen (1999) were able to
show that this replacement plausibly leads to a
mass function almost identical to that which Sheth \& Tormen (1998)
had earlier fitted to a subset of the numerical 
data we analyse below. Note that at present there is no good numerical
test of analytic predictions for the low mass tail of the mass function,
and our analysis in this paper does not remedy this situation.

Several authors have considered halo mass functions in models with
scale-free Gaussian initial fluctuations. The attraction is that 
clustering should be self-similar in time in such models when
$\Omega=1$. This expectation is easy to test numerically and
appears substantiated by the available simulation data (Efstathiou 
\etal\ 1988, Lacey \& Cole 1994).  Deviations from self-similarity 
can be used to isolate numerical artifacts which break the scaling. 
The CDM power spectrum is not scale-free although the variation of the
effective spectral index with wavenumber is gentle. Recently, a number
of very large CDM simulations have been performed by Gross \etal\
(1998) and Governato \etal\ (1999). These confirmed the deviations 
from the P-S prediction found in earlier work and extended coverage
to both higher and lower masses.  An interesting question is
how the non-power-law nature of the fluctuation spectrum 
affects the mass function; P-S theory predicts there
to be no effect when the mass function is expressed in 
suitable variables.  However, Governato \etal\ (1999) find evidence
that the high-mass deviation from the P-S prediction increases
with increasing redshift, although the effect is small.

The essence of group finding is to convert a discrete representation
of a continuous density field into a countable set of ``objects''. In
general, this conversion is affected both by the degree of
discreteness in the realisation (the particle mass) and by the
detailed characteristics of the object definition algorithm. The
definition of object boundaries is somewhat arbitrary, and one can 
expect the characteristics of the object set to vary according to the
specific assumptions adopted. This introduces uncertainties when
comparing simulations and analytic models. We examine here two of 
the standard algorithms used in earlier literature, the 
friends-of-friends and spherical overdensity group-finders (Davis 
\etal\ 1985, Lacey \& Cole 1994).  Motivated by spherical 
collapse models, both attempt to identify virialized regions that 
are overdense by a factor $\sim 200$ with respect to the global mean.
Some comparisons of these algorithms have already been published by
Lacey \& Cole (1994) who found small but measurable differences.

In reality, halos possess a variety of physical characteristics that can be
employed by observers to form new, and possibly different, countable
sets. For example, the Coma cluster is a single object to an X-ray
observer but contains hundreds of visible galaxies. Similarly, very
high resolution simulations reveal a myriad of smaller,
self-bound ``sub-halos'' within the virial region of each parent halo
(e.g. Moore \etal\ 1999, Klypin \etal\ 1999). The distinction between
the larger halo and its substructure is largely one of density; the
sub-halos are bounded at a much higher density 
than the parent object.  Such subtleties further
complicate comparisons between theory, experiment and observation, and
in all such analysis it is important to take careful account of
the specific algorithms used to define the objects considered.

In Section~\ref{sim_det}, we describe the simulations and the two halo
finders that we employ when determining halo mass functions. In
Section~\ref{consist}, we investigate the consistency of the mass
functions derived from different simulations of the \tcdm\ and \lcdm\
cosmologies. In Section~\ref{comps}, we present our results for the
\tcdm\ and \lcdm\ models and compare them with P-S theory and with the 
Sheth-Tormen fitting formula. In Section~\ref{gen_mfn}, we examine the
evolution of the high mass end of the mass function with redshift. Using
the insight that this provides, we generalize our results and show that if
a single linking length is used to define halos, then a single fitting
formula, very close to that proposed by Sheth \& Tormen, accurately 
describes the mass functions in \tcdm\, \lcdm\, SCDM and OCDM models over
a wide range of redshifts. We present and discuss our conclusions in
Section~\ref{cc}, where we also explain how to obtain some of the 
simulation data and the analysis software used in our study. Appendix 
A tests the influence of numerical parameters on our measured mass 
functions, while appendix B gives ``best fit'' analytic representations
of a number of our mass functions.

\begin{table*}
\centering
\begin{minipage}{140mm}
\caption{N-body simulation parameters.}
\begin{tabular}{@{}lrrlrrrrr@{}}
Run  &  $\Omega_0$ & $\Lambda_0$ & $\Gamma$ & $\sigma_8$ & 
$N_{\rm part}\phantom{aaa}$ 
     & $L_{\rm box}/h^{-1}{\rm Mpc}$ & $m_{\rm particle}/h^{-1}\msun$ & 
$r_{\rm soft}/h^{-1}{\rm kpc}$\\
\tcdm-gif &   1.0  & 0.0 & 0.21 & 0.60 & $16\,777\,216$ & $84.5\phantom{bbbb}$ 
& $1.00\times10^{10}\phantom{ccc}$ & $30\phantom{aaaaaa}$\\
\tcdm-int &   1.0  & 0.0 & 0.21 & 0.60 & $16\,777\,216$ & $160.3\phantom{bbbb}$ 
& $6.82\times10^{10}\phantom{ccc}$& $10\phantom{aaaaaa}$\\
\tcdm-512 &   1.0  & 0.0 & 0.21 & 0.51 & $134\,217\,728$ & $320.7\phantom{bbbb}$ 
& $6.82\times10^{10}\phantom{ccc}$& $30\phantom{aaaaaa}$\\
\tcdm-virgo &   1.0  & 0.0 & 0.21 & 0.60 & $16\,777\,216$ & $239.5\phantom{bbbb}$ 
& $2.27\times10^{11}\phantom{ccc}$& $30\phantom{aaaaaa}$\\
\tcdm-hub &   1.0  & 0.0 & 0.21 & 0.60 & $1\;000\,000\,000$ & $2000.0\phantom{bbbb}$ 
& $2.22\times10^{12}\phantom{ccc}$& $100\phantom{aaaaaa}$\\
\lcdm-gif & 0.3  & 0.7 & 0.21& 0.90 & $16\,777\,216$ & 
$141.3\phantom{bbbb}$ & $1.40\times10^{10}\phantom{ccc}$& $25\phantom{aaaaaa}$\\
\lcdm-512 &   0.3  & 0.7 & 0.21 & 0.90 & $134\,217\,728$ & $479.0\phantom{bbbb}$ 
& $6.82\times10^{10}\phantom{ccc}$& $30\phantom{aaaaaa}$\\
\lcdm-hub & 0.3  & 0.7 & 0.21& 0.90 & $1\,000\,000\,000$ & 
$3000.0\phantom{bbbb}$ & $2.25\times10^{12}\phantom{ccc}$& $100\phantom{aaaaaa}$\\
 &  & & & & & & &\\
\tcdm-test1 &   1.0  & 0.0 & 0.21 & 0.60 & $1\,000\,000$ & $200.0\phantom{bbbb}$ 
& $2.22\times10^{12}\phantom{ccc}$& $100\phantom{aaaaaa}$\\
\tcdm-test2 &   1.0  & 0.0 & 0.21 & 0.60 & $1\,000\,000$ & $200.0\phantom{bbbb}$ 
& $2.22\times10^{12}\phantom{ccc}$& $30\phantom{aaaaaa}$\\
\tcdm-test3 &   1.0  & 0.0 & 0.21 & 0.60 & $8\,000\,000$ & $200.0\phantom{bbbb}$ 
& $2.77\times10^{11}\phantom{ccc}$& $100\phantom{aaaaaa}$\\
\end{tabular}
\end{minipage}
\end{table*}

\section{Simulation details}\label{sim_det}

\subsection{The simulations}\label{the_sims}

For the main analysis of this paper we measure halo mass functions 
in a number of N-body simulations carried out by the Virgo consortium 
for two cosmological models, \tcdm\ and \lcdm, as introduced by Jenkins
\etal\ (1998). The simulation parameters are listed in
Table~1.  The largest calculations are the two ``Hubble volume"
simulations, each with $10^9$ particles in boxes of volume $8$ and
$27h^{-3}{\rm Gpc}^3$\footnote{We express the Hubble constant as
$H_0=100h {\rm km s}^{-1}{\rm Mpc}^{-1}$.} respectively. These
simulations allow the mass function to be determined for very massive
clusters ($>10^{15}\hmsun$) with relatively small Poisson errors.
In addition, we have analysed several simulations of smaller volumes,
but with better mass resolution, from which the mass function can be
determined reliably down to masses of a few times $10^{11}\msun$.
In section 5 we also include data from the large SCDM simulations
carried out by Governato \etal\ (1999) and from the simulations of
SCDM and OCDM discussed in Jenkins \etal\ (1998).

The models listed in Table~1 are all normalised so as to be consistent with
the observed local abundance of rich galaxy clusters (White, Efstathiou \&
Frenk 1993, Eke, Cole \& Frenk 1996, Viana \& Liddle 1996) and are also
consistent with the standard COBE normalization (e.g. Ratra \etal\
1997). However, for the most part, the precise normalisation is unimportant
for the purposes of this paper.  The power spectrum of the initial
conditions of all the simulations except \lcdm-hub \&
\lcdm-512 was set up using the transfer function given by Bond \&
Efstathiou (1984),
\begin{equation}\label{powspec}
 T(k) = {1\over\bigg[1+\big[aq + (bq)^{3/2} 
+ (cq)^2\big]^{\nu}\bigg]^{1/\nu}}, 
\end{equation}
where $q = k/\Gamma$, $a = 6.4\hmpc$, $b = 3\hmpc$, $c = 1.7\hmpc$ and
$\nu=1.13$. For the \lcdm-hub and \lcdm-512 simulations, the transfer
function was computed using CMBFAST (Seljak \& Zaldarriaga 1996),
assuming $h=0.7$ and $\Omega_{\rm baryon}h^2 =0.0196$ (Burles and
Tytler 1998). In all models, the slope of the primordial power
spectrum was taken to be unity. Full details of how the simulations
were set up are given in Jenkins \etal\ (1998), for simulations ending in
-gif -int -virgo and -test, and in Evrard \etal\ (2000), for 
simulations ending in -hub. More recently, we have completed 
$512^3$-particle simulations of the \tcdm\ and \lcdm\
models using essentially the same code that we used for the Hubble
volume calculations.

Each simulation yields a determination of the mass function over a mass
range dictated by the particle mass and the number of particles in the
simulation. To aid understanding, we have performed a set of smaller test
simulations (see bottom of Table~1), designed to investigate the
sensitivity of the mass function to changes in a single numerical parameter
at a time. The parameters that we vary are the particle mass, the
gravitational softening and the starting redshift.  These checks 
(discussed in appendix A) also give
an indication of the number of particles required to determine the mass
function satisfactorily using different halo finders.

\subsection{Halo finders}\label{halo_finders}

We use two different algorithms to identify dark matter halos: the
friends-of-friends (FOF) algorithm of Davis \etal\ (1985), and the
spherical overdensity (SO) finder described by Lacey \& Cole
(1994). The FOF halo finder depends on just one parameter, $b$, which
defines the linking length as $bn^{-1/3}$ where $n$ is the mean
particle density. An attractive feature of the FOF method is that
it does not impose any fixed shape on the halos.  A disadvantage is
that it may occasionally link two halos accidentally through a chance
bridge of particles.  In the limit of very large numbers of particles
per object, FOF approximately selects the matter enclosed
by an isodensity contour at $1/b^3$. 

In the SO algorithm, the mass of a halo is evaluated in a spherical
region. There is one free parameter, the mean overdensity, $\kappa$,
of the halos. There are many possible ways of centering the spherical
region. In our case, the centre is determined iteratively, after
making an initial guess based on an estimate of the local density for
each particle, re-centering on the centre-of-mass, growing a sphere
outwards about the new centre until it reaches the desired mean
overdensity, and recomputing the centre-of-mass. After several
iterations, the motion of the centre becomes small.  An alternative
method consists of centering the sphere on the local maximum enclosed
mass at fixed overdensity, but this is more computationally intensive
-- a strong disincentive when identifying SO groups in the Hubble
volume simulations.

The conventional choices for $\Omega=1$ cosmologies are FOF($b=0.2$)
and SO($\kappa$=180) (Davis \etal\ 1985; Lacey \& Cole\ 1994). For
models where $\Omega\ne1$, the choice is less obvious.  At late times,
groups stop growing and the appropriate linking length or bounding
density should plausibly become constant in physical coordinates. At
early times $\Omega\simeq 1$, and it is the corresponding comoving
quantities that are most naturally kept fixed.  One needs to decide
how to make the transition between these two regimes. Lacey \& Cole
(1994) and Eke, Cole \& Frenk (1996) have done this using the
spherical top-hat collapse model. We adopt their approach for the
analysis in Section~4, taking FOF($b=0.164$) and SO($\kappa=324$)
for \lcdm\ at $z=0$. This choice cannot be rigorously justified,
however, and we find in Section~5 that simpler results are obtained 
by using the conventional Einstein-de Sitter parameters also for this
cosmology and for OCDM.

\figstart{figure=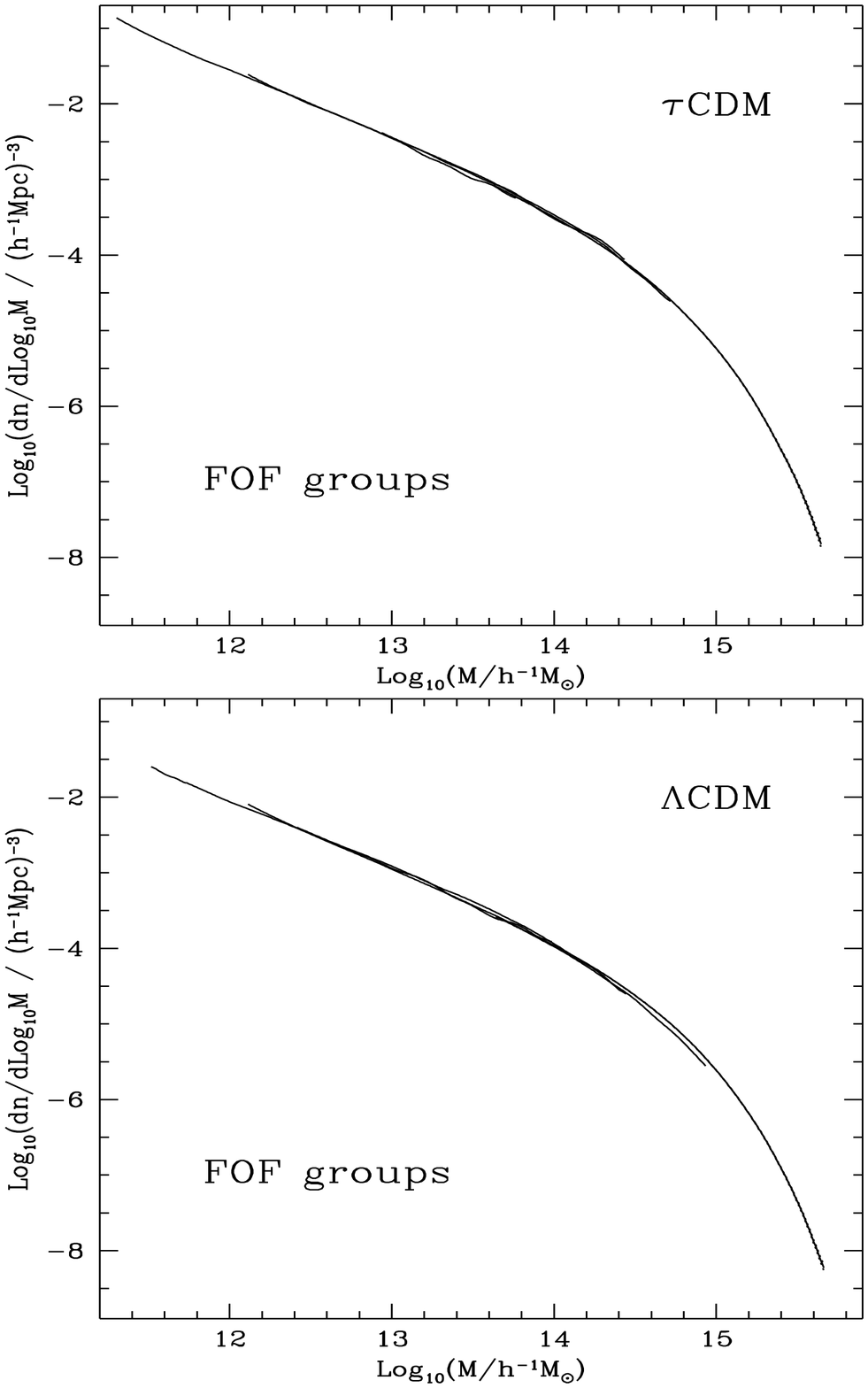,width=8.0cm}
\label{fofcomb}
\caption{ Friends-of-friends differential mass functions for dark
matter halos in the \tcdm\ and \lcdm\ simulations. Halos were
identified using linking lengths of 0.2 and 0.164 respectively.  The
different curves correspond to the various simulations detailed in
Table~1. The mass resolution in the simulations varies by more than
two orders of magnitude and the volume surveyed by more than four
orders of magnitude. In all cases, the mass functions are truncated 
at the low mass end at the mass corresponding to 20 particles, and at
the high mass end at the point where the predicted Poisson abundance
errors reach 10\%. The simulations match up very well.}
\figend

\figstart{figure=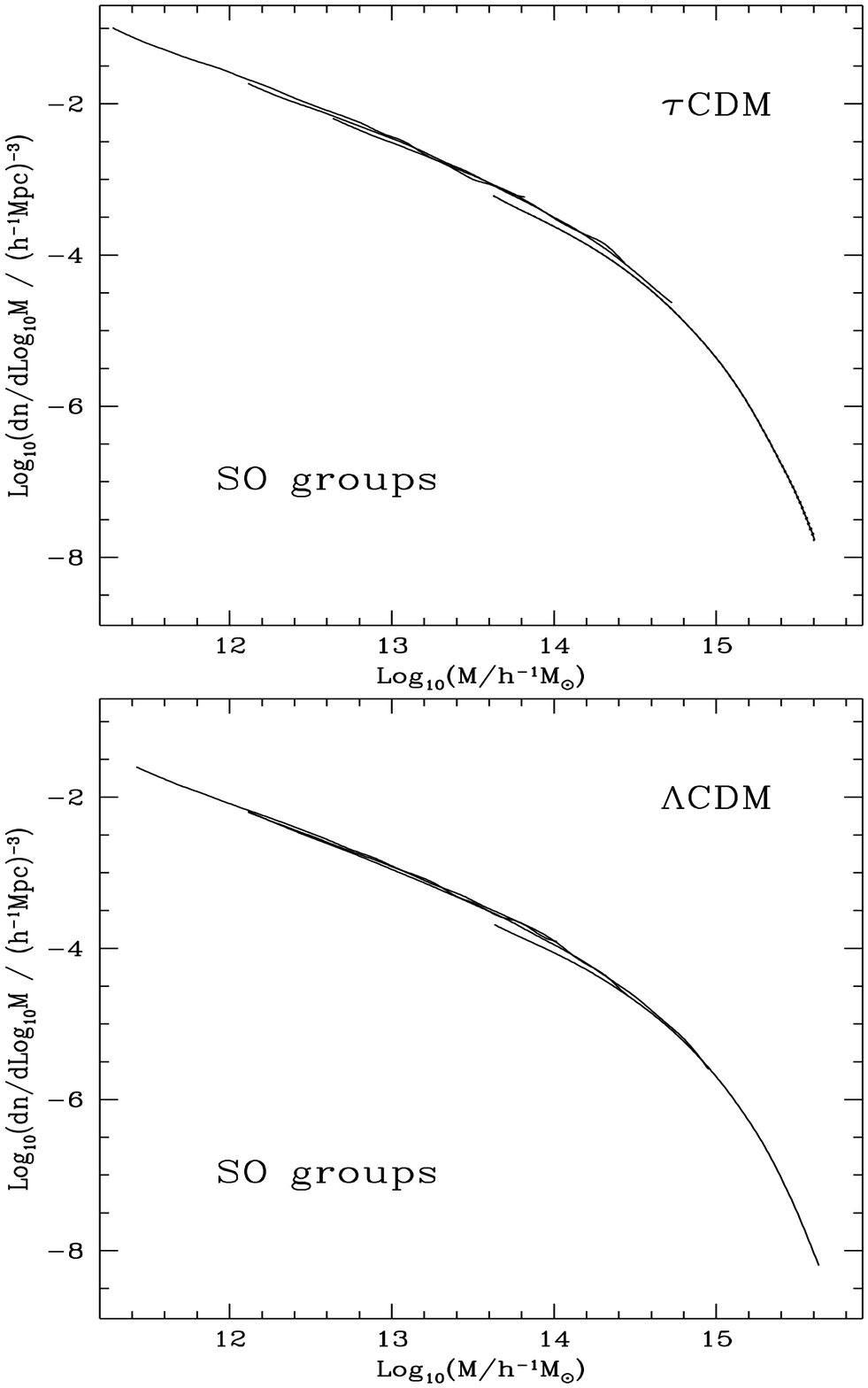,width=8.0cm}
\label{socomb}
\caption{Spherical overdensity differential mass function for dark
halos in the \tcdm\ and \lcdm\ simulations. Halo masses were defined
at mean interior overdensities of 180 and 324 respectively. The
different curves correspond to the various simulations detailed in
Table~1 and are truncated as in Fig.1. The 
simulations do not match up as well here as in Fig.1.  Simulations with
coarse mass resolution seem to underestimate the halo abundance near
their lower mass limit. See the text for further details.}
\figend

\section{Consistency checks}\label{consist}

It is important to check the reliability of mass
functions determined from simulations. From a formal point of view
this is impossible, since there is no known analytic solution
for any realistic halo definition. It is, however, possible to check 
for consistency between different simulations and different halo
definitions. The mass ranges covered by the different simulations
in each of our sets overlap considerably. As we show
below, the agreement in these overlap regions is far from perfect.
One source of difference is just the Poisson error due to the
finite number of clusters in each simulation. In the following we
will only use numerical data for which such Poisson errors
are below 10\% and are negligible compared to systematic errors.
These, we attribute to weak dependences of the measured mass functions
on some of the numerical parameters of our simulations. Such
systematics appear to be smaller than about 10\% for the FOF 
halo-finder and about 30\% for the SO halo-finder, and are therefore
too small to be a major concern. It seems unlikely that 
observational determinations of halo mass functions will reduce
systematic uncertainties to such small levels in the foreseeable 
future.

Defining $n(M)$ to be the number of halos with mass less than $M$, we
plot the differential mass functions, ${\rm d}n/{\rm d}\ln M$ for all
\tcdm\ simulations in Table~1 with $\sigma_8=0.6$ and for all \lcdm\
simulations with $\sigma_8=0.9$, excluding the last three test
simulations.  The mass functions for the test simulations are plotted
in Appendix~A and compared with other simulations to show how much the
mass function can vary as a result of changes in the particle mass,
gravitational softening and starting redshift.

The FOF mass functions for our two cosmologies are displayed
in Fig.~1. The numerical data have been smoothed with a kernel 
which is gaussian in
$\log_{10}(M)$ with RMS width 0.08. This
smoothing erases high frequency Poisson noise and provides a 
continuous curve for comparison with analytic predictions. Using
a `bin' shape without sharp edges also reduces fluctuations at the low
mass end where the halo masses are integer multiples of the particle
mass.  For a power-law mass function, $F\equiv {\rm d}n/{\rm
d}M\propto M^{-\alpha}$, such smoothing raises
the amplitude by approximately $\exp(\alpha^2/59)$. For
the \tcdm\ mass function, this factor is about 1.03 at
$10^{14}\hmsun$, 1.14 at $10^{15}\hmsun$, and 1.7 at the highest
masses we plot; similar numbers apply to \lcdm.  Poisson statistics
lead to RMS uncertainties in the smoothed curves which are
\begin{equation}
  {\delta F_{\rm RMS}\over F} \simeq 1.88 \Big(V_{\rm sim}{{\rm
  d}n\over{\rm d}\log_{10}M}\Big)^{-1/2}\exp(-\alpha^2/79),
\end{equation}
where $V_{\rm sim}$ is the volume of the simulation considered
and errors are correlated over a distance comparable to
the smoothing length-scale.  As noted above, we only plot curves for
which this uncertainty is below 10\% and is small compared to
other sources of error. 

As can be seen in Fig.~1, FOF mass functions for different simulations
match well even when the number of particles per halo is
small. The linking parameter was taken to be $b=0.2$ for \tcdm\ and
$b=0.164$ for \lcdm. For the smaller boxes, Poisson fluctuations are clearly
visible at the high mass end of the plotted curves. Such
fluctuations are much less pronounced in the curves derived from the
Hubble volume simulations, because the mass function is then 
much steeper at the point where discreteness noise becomes appreciable
and our smoothing erases features more efficiently. 

Fig.~2 shows a similar comparison of mass functions obtained using the SO
halo-finder. The mean overdensity was set to 180 and 324 in the \tcdm\ and
\lcdm\ cases respectively. Here we see a systematic effect: the abundance
at given halo mass in a simulation with large particle mass is always lower
than that found in a simulation with smaller particle mass. The difference
is particularly pronounced between the Hubble volume simulations and the
others, but this merely continues a trend of increasing discrepancy as the
halo mass at the matching point becomes larger. The tests carried out in
Appendix~A show that these mismatches result from resolution effects on
halo structure which particularly affect halos identified with the SO
algorithm. Robust results require a minimum of about 100 particles per SO
halo, but only a minimum of about 20 particles per FOF halo.

\section{Comparison with analytic models}\label{comps}

In this section, we compare some popular analytic models to the
mass functions constructed above. It proves convenient 
to use the quantity $\ln\sigma^{-1}(M,z)$ as the mass variable instead
of $M$, where $\sigma^2(M,z)$ is the variance of the linear density 
field, extrapolated to the redshift $z$ at which halos are identified,
after smoothing with a spherical top-hat filter which encloses mass 
$M$ in the mean. This variance can be expressed in terms of
the power spectrum $P(k)$ of the linear density field extrapolated to
redshift zero as:
\begin{equation}\label{vardeff}
\sigma^2(M,z)  =  {b^2(z)\over2\pi^2}\int_0^\infty k^2P(k)W^2(k;M){\rm d}k,
\end{equation} 
where $b(z)$ is the growth factor of linear perturbations normalised
so that $b=1$ at $z=0$, and $W(k;M)$ is the
Fourier-space representation of a real-space top-hat filter
enclosing mass $M$ at the mean density of the universe.

We define the mass function $f(\sigma, z;X)$ through:
\begin{equation}\label{deff}
   f(\sigma, z;X) \equiv \frac{M}{\rho_0}{{\rm d}n_{X}(M, z)\over{\rm d}\ln\sigma^{-1}},
\end{equation}
where $X$ is a label identifying the cosmological model and halo
finder under consideration, $n(M, z)$ is the abundance of halos with 
mass less than $M$ at redshift $z$, and $\rho_0(z)$ is the mean density 
of the universe at that time.

The advantage of using $\ln\sigma^{-1}$ as the mass variable is most
evident when we consider the analytic models to which we compare. As
we will see below, the Press-Schechter model predicts a simple
analytic
form for $f(\sigma, z)$ which has no explicit dependence on redshift,
power spectrum or cosmological parameters; a single mass function
describes structure in all gaussian hierarchical clustering models
at all times in any cosmology provided abundances are plotted in
the $f - \ln (\delta_c/\sigma)$ plane, where $\delta_c$ is a
threshold parameter, possibly dependent on $\Omega$, which we discuss
later. This very simple structure carries over to extensions of the
Press-Schechter analysis such as that presented by Sheth, Mo \&
Tormen (1999). For a power-law linear power spectrum and $\Omega=1$, 
clustering is expected to be self-similar in time on general grounds
independent of the P-S model (e.g. Efstathiou \etal\ 1988, Lacey 
\& Cole\ 1994), implying again that mass functions at different times
should map onto a unique curve in the $f-\ln\sigma^{-1}$\ plane
(although this curve could be a function of the power-law spectral index).
For CDM power spectra, the spectral index varies quite weakly with scale so
one might expect at most a weak dependence on redshift in this
plane. At worst, use of $\ln\sigma^{-1}(M)$ as the ``mass'' variable
``factors out'' most of the difference in the mass functions between
different epochs, cosmologies and power spectra, and so
allows a wider comparison, both among our own simulations
and between these and those by other authors.

In Subsection~4.1, we describe the two analytic models with which we 
compare explicitly to our numerical results in subsection~4.2. 
 
\subsection{Analytic models}\label{ps_mod}
 
The Press-Schechter model (e.g. Press \& Schechter\ 1974, Bond {\it et
al.}\ 1991, Lacey \& Cole\ 1993) predicts a mass-function given by:
\begin{equation}\label{ps_massfn}
   f(\sigma; {\rm P{\rm-}S}) = \sqrt{2\over\pi}
   {\delta_c\over\sigma}\exp\bigg[-{\delta_c^2\over2\sigma^2}\bigg],
\end{equation}
where $\delta_c$ is a threshold parameter usually taken to be the
extrapolated linear overdensity of a spherical perturbation at the
time it collapses. In an Einstein-de Sitter cosmology
$\delta_c=1.686$. 
In other cosmologies $\delta_c$ is sometimes assumed to be a 
weak function of $\Omega$ and $\Lambda$ (see e.g. Eke, Cole \& Frenk 1996). 
The P-S mass function has the normalisation property:

\begin{equation}\label{integcond}
   \int_{-\infty}^\infty f(\sigma; {\rm P{\rm-}S}){\rm d}\ln\sigma^{-1} = 1.
\end{equation} 
This implies that all of the dark matter is attached to halos of some
mass.

Sheth \& Tormen (1999, hereafter S-T) have introduced a new formula for the
mass function (see their eqn.~10) which, for empirically determined choices
of two parameters, gives a good fit to the mass functions measured in a
subset of the N-body simulations analysed in this paper (the
simulations ending in -gif in Tables~1 and~2).
Their mass function can be expressed as:
\begin{equation}\label{sheth_tormen}
  f(\sigma; {\rm S{\rm-}T}) = A\sqrt{{2a\over\pi}}
\bigg[1+\big({\sigma^2\over a\delta_c^2}\big)^p\bigg]
   {\delta_c\over\sigma}\exp\bigg[-{a\delta_c^2\over2\sigma^2}\bigg],
\end{equation}
where A=0.3222, $a=0.707$ and $p=0.3$. Sheth, Mo and Tormen (1999,
SMT) extended the excursion set derivation of the P-S formula
developed by Bond \etal\ (1991) to include an approximate treatment of
ellipsoidal collapse, and showed this to produce a mass function
almost identical in shape to eqn.(\ref{sheth_tormen}). Their
derivation forces this mass function to obey the integral relation of
eqn.~(\ref{integcond}). Below, we compare their mass function with our
N-body results, including mass scales, redshifts and cosmologies other
than those it was originally forced to fit. In a separate paper
(Colberg et al 2000), we compare the clustering of halos in our Hubble
Volume simulations with the predictions of the SMT model.

Other analytic models for the halo mass function have been proposed 
recently by Monaco (1997a,b) and Lee \& Shandarin (1998). These are
substantially poorer fits to our data than the Sheth, Mo \& Tormen
(1999) model and we do not consider them further in this paper. Other
published discussions of ``Press-Schechter'' predictions have made use
of filter functions other than the spherical top hat, or have treated the
threshold $\delta_c$ as an adjustable parameter. We will not pursue
the first of these possibilities at all, and we only deviate from
standard assumptions about $\delta_c$ in Section (\ref{gen_mfn}).
There we show that taking $\delta_c=1.686$ in {\it all} cosmologies
leads to excellent agreement with our numerical data if
halos are defined at fixed overdensity independent of $\Omega$,
rather than at an $\Omega$-dependent overdensity as in the more 
conventional approach.

\subsection{Comparison to simulated mass functions}\label{fit_res}

In order to make proper comparisons between analytic models and our 
simulated halo mass functions, we smooth the analytic predictions in the
same way as we smoothed the simulation results in Section~3.  In practice,
this smoothing changes the predictions very little; the difference is
only perceptible at high masses where the curves are very steep.

Interpolation formulae that accurately represent our (unsmoothed) mass
functions are given in Appendix~B and plotted (after smoothing) in 
Figs.~3 and~ 4.  Fig.~3 compares the FOF(0.2) halo mass function
in the \tcdm\ simulations with the P-S and S-T formulae assuming
$\delta_c=1.686$. At the high mass
end ($\ln\sigma^{-1} > 0$), the simulation results lie well above the P-S
prediction.  This discrepancy has been observed by a large number of
authors (e.g. Efstathiou \etal\ 1988, White, Efstathiou \& Frenk 1993,
Gross \etal\ 1998, Governato \etal\ 1999).  Our simulations confirm that
the divergence increases at even larger halo masses than those accessible
to previous simulators.  For $\ln\sigma^{-1} <
0.3$, the P-S curve overestimates the simulated mass functions, and
at the characteristic mass $M_*$ (where $\sigma=\delta_c$ and so
$\ln\sigma^{-1} = -0.52$), the halo abundance is only 60\% of
the P-S prediction. This conclusion agrees with the results of 
Efstathiou \etal\ (1988) and Gross \etal\
(1998) who found a similar discrepancy for a number of different
cosmological models. Adjusting the simulated mass functions by altering
halo-finder parameters, or the analytic predictions by altering
$\delta_c$, tends to shift the relevant curves in the $\ln f  - 
\ln\sigma^{-1}$ plane without much 
altering their shape. As a result, it does not significantly improve
the overall agreement between the P-S model and the numerical data.

By contrast, the S-T mass functions give an excellent fit to the 
N-body results in Fig.~3.  Good agreement is to be
expected at the low mass end since a subset of our simulation data 
was used by Sheth \& Tormen (1999) to determine the parameters of
their fitting function. This fitting function matches our numerical
data for \tcdm-FOF over
their entire range, including large masses which were not considered in
the original fit.

\figstart{figure=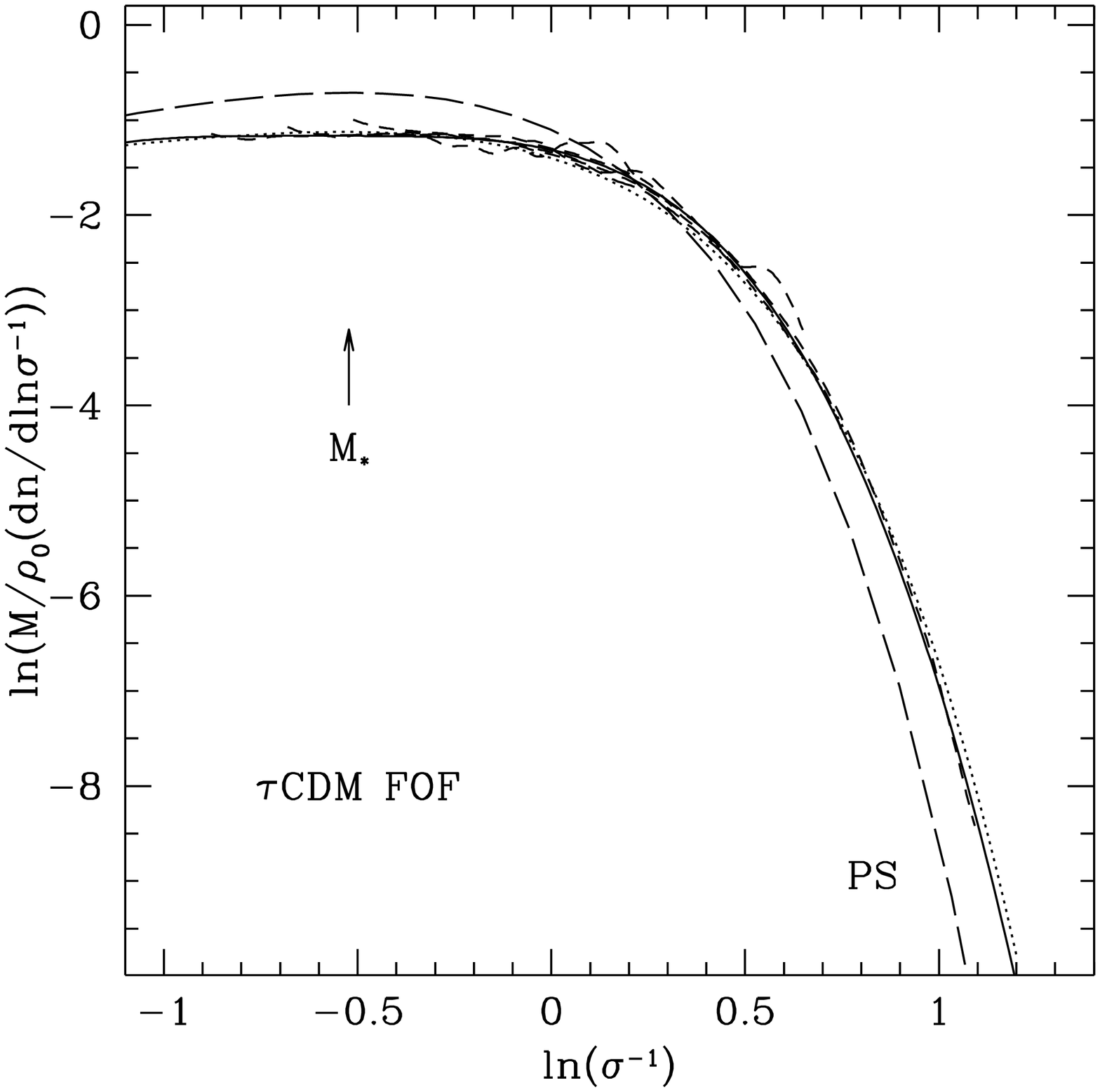,width=8.0cm}
\label{figure-3}
\caption{A comparison of analytic models with the halo mass function 
at $z=0$ in our N-body simulations of the \tcdm\ cosmology. Halos were found
using the FOF algorithm with $b=0.2$. The short dashed lines show
results from the individual \tcdm\ simulations used in this paper, 
the solid curve is the fit of eqn.~(\ref{tcdm_fit}) to the combined 
results of the simulations, while the dashed line shows the P-S 
prediction and the dotted line, the S-T prediction, both using 
$\delta_c=1.686$.  The arrow marks the characteristic mass scale, 
$M_*$, where $\sigma(M_*)=\delta_c$, and corresponds to the 
position of the peak in the Press-Schechter mass multiplicity
function. Note that we use natural logarithms in this plot.}
\figend

\figstart{figure=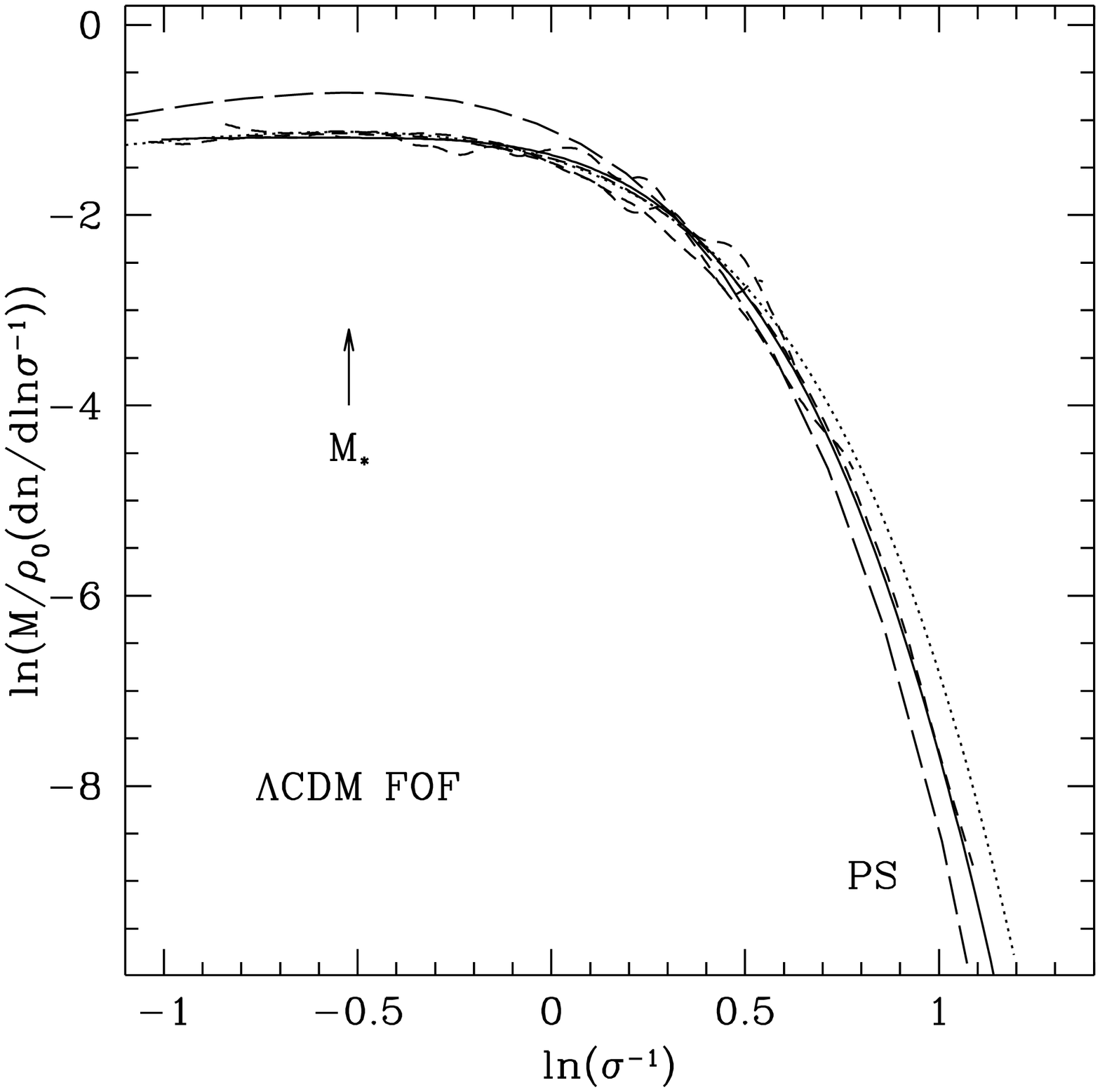,width=8.0cm}
\label{figure-4}
\caption{A comparison of analytic models with the halo mass function
at $z=0$ in 
our N-body simulations of the \lcdm\ cosmology. Halos were found
using the FOF algorithm with $b=0.164$. The short dashed lines 
show results from the individual \lcdm\ simulations used in this 
paper, the solid curve is the fit of eqn.~(\ref{lcdm_fit}) to the 
combined results of the simulations, while the long dashed line 
shows the P-S prediction and the dotted line, the S-T prediction,
both using $\delta_c=1.675$. The arrow marks the characteristic 
mass scale, $M_*$, where $\sigma(M_*)=\delta_c$, and corresponds 
to the position of the peak in the Press-Schechter mass function.
Note that we use natural logarithms in this plot.}
\figend

Fig.~4 shows the analogous comparison in the case of the \lcdm\ model.
Here, the FOF(0.164) halo mass function in the simulations is compared to
the P-S and S-T predictions using $\delta_c=1.675$ as advocated by
Eke \etal\ (1996) for this cosmology. 
With these choices of parameters, the P-S curve gives a better
fit to the simulated mass function at high masses than in the
\tcdm\ case, although it still underestimates the abundance
of high mass clusters and substantially overestimates the
abundance near $M_*$. The S-T model is a poorer
fit to the numerical data than for \tcdm\ , substantially
overestimating abundances at high masses. However, as we will show in the
next section, this disagreement is all but removed by different
(and simpler) assumptions about the appropriate parameters for
halo-finders when $\Omega < 1$.

All the above comparisons refer to halos identified using the FOF 
halo-finder with standard parameters. We have checked that very 
similar results are obtained if halos are identified using the SO 
halo-finder, again with standard parameters. This similarity was
previously noted by Tormen (1998). 

It is interesting to compute the fraction of the total cosmic
mass density which is included in halos over the full range of
validity of our simulation mass functions ($\sim 3\times 10^{11} 
\ {\rm to}\ \sim 5\times10^{15} \hmsun$). Using our fits to the 
FOF data in Figs~3 and 4, we find this fraction to be 0.37 both in
\tcdm\ and in \lcdm\ . The corresponding fraction for the P-S mass
function is 0.50. Where is the remaining mass?  Clearly, higher
resolution simulations will show some proportion to be in halos too small
to have been resolved by our current simulations. However, there is
no guarantee that such higher resolution simulations will produce
a total halo mass fraction which converges to unity. Much
of the dark matter may not lie in halos identified by an FOF(0.2) 
(or other) halo-finder, but rather in a smooth, low-density 
component, perhaps in extensions of the identified halos
beyond their artificial $b=0.2$ boundaries. This possibility is suggested 
by the fact that our simulated mass functions remain low compared
to the P-S curve at the smallest masses for which we can measure
them. A straightforward extrapolation of our FOF(0.2) curves
contains only $\sim70\%$ of the total mass. Unfortunately, this 
extrapolation is not unique as shown by the fact that the S-T
mass function can fit all our numerical data and yet is normalised
to give a total halo mass fraction of unity. More numerical
data are needed to study the low mass behaviour of the mass function
in order to resolve this issue.

\section{Towards a general fitting formula}\label{gen_mfn}

In this section, we first show that, when expressed in terms of
$\ln\sigma^{-1}$, the mass functions in our two cosmological models
vary only very little with redshift, or equivalently, with the effective
slope of the power spectrum. With our current definition of halos, however,
the mass functions in the \tcdm\ and \lcdm\ models are different. In
subsection~5.2, we show that this difference all but disappears if, instead
of using a FOF linking length that varies with $\Omega$, we simply identify
clusters with a constant linking length, $b=0.2$, in both
cosmologies. A general fitting formula can then
accurately describe the halo mass function in a wide range of
cosmological models.

\subsection{Comparison of the mass function at different redshifts} 
\label{evol_mfn}
For a scale-free power spectrum, the mass function expressed in terms of
$\ln\sigma^{-1}$ should be independent of redshift.  Any differences must be
due to Poisson sampling or to systematic errors introduced by numerical
inaccuracies that break the scaling laws. For a CDM spectrum, for which the
spectral slope is a function of scale, there could, in principle, be
genuine evolution of the mass function but, if the amount of evolution is
small, it may be masked by numerical effects.

In the \tcdm\ model, a suitable rescaling of the length and mass units
allows the mass function at non-zero redshift to be regarded as the
redshift zero mass function of a simulation with a different power spectrum
but identical normalisation. Differences in the $f-\ln\sigma^{-1}$ plane
between mass functions at different redshifts can therefore indicate a
dependency of the mass function either on the shape of the power spectrum or
on numerical effects.  In order to exploit this regularity, and also
to compare our results with those of Governato \etal\ (1999), we
parametrise each simulation output by an effective power spectral
slope $n_{\rm eff}$,
\begin{equation}
      n_{\rm eff} = 6 {{\rm d}\ln\sigma^{-1}\over{\rm d}\ln
      M\phantom{+}} -3,
\label{defnstar}\end{equation}
where we evaluate the derivative at the point where $\sigma=0.5$,
corresponding roughly to cluster scales for $\Omega=1$ and $z=0$. For a 
scale-free power spectrum, $n_{\rm eff}$ is the power-law index in
$P(k)\propto k^n$.  We have calculated the FOF(0.2)
mass function for the \tcdm-hub simulation at redshifts $z = 0.0, 0.18,
0.44$ and $0.78$, for which $n_{\rm eff} = -1.39, -1.48, -1.58$ and -1.70
respectively.  We can extend this range of spectral slopes by considering,
in addition, the mass functions determined by Governato \etal\ (1999). 
Four outputs of their SCDM model correspond to $\sigma_8 = 1.0, 0.7, 0.47$
and 0.35, for which $n_{\rm eff} = -0.71, -0.90, -1.12$ and -1.28.
respectively. Thus, the two sets of simulations cover the range -0.71 to
-1.7 in spectral slope but do not overlap.

\figstart{figure=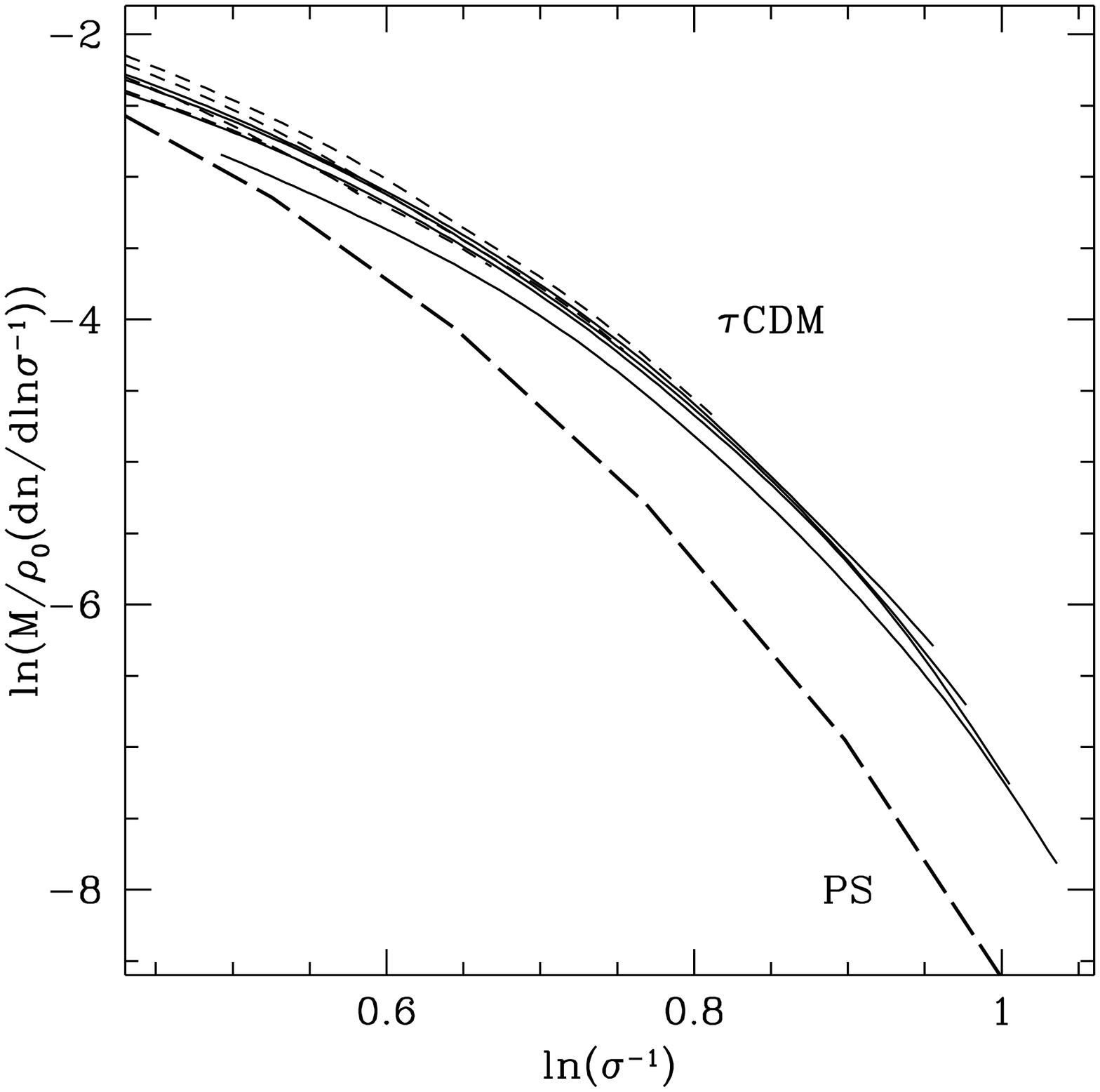,width=8.0cm}
\label{figure-5}
\caption{A comparison of mass functions in different simulations at
different epochs. The full curves show the FOF(0.2) mass functions for the
\tcdm-hub simulation at redshifts $z=0, 0.18, 0.44$ and 0.78. The dashed
curves show the corresponding functions for the SCDM simulations of
Governato et al (1999) at four epochs for which $\sigma_8=$1.0,0.7,0.47
and 0.35 respectively. The heavy dashed curve shows the P-S model
function. Both simulation datasets show a weak trend with $\sigma_8$, but the
trends are opposite! See the text for discussion.  Note that we use
natural logarithms in this plot.}  \figend

\figstart{figure=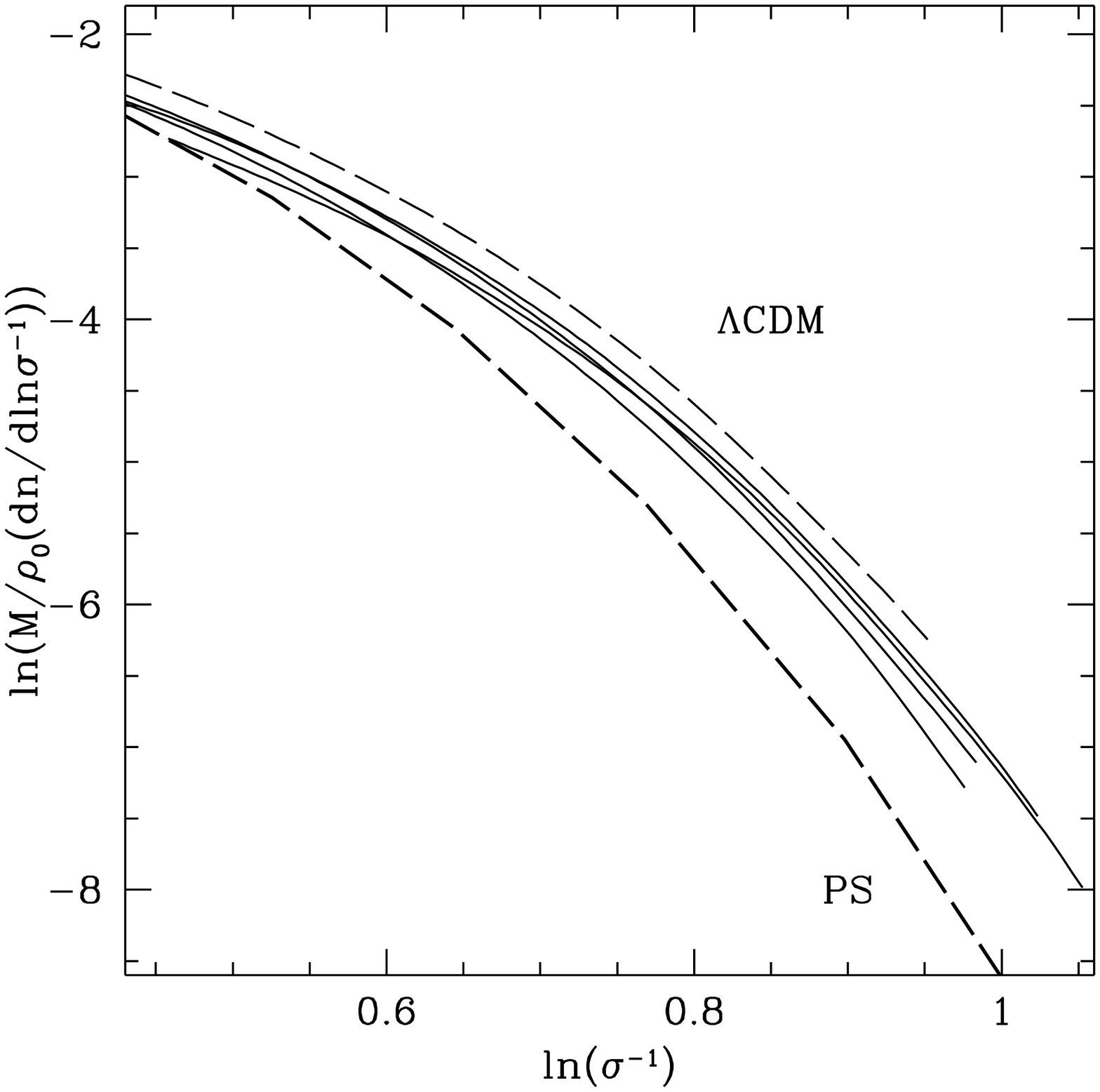,width=8.0cm}
\label{figure-6}
\caption{A comparison of mass functions in different simulations at
different epochs. The full curves show the FOF mass function for the
\lcdm-hub simulation at redshifts $z=0, 0.27,0.96$ and $1.44$. The
first three form a sequence going from bottom to top, while the
$z=1.44$ output is slightly lower than $z=0.96$. The light dashed
curve shows the \tcdm-hub ($z=0$) mass function.  The heavy dashed
curve shows the P-S theory model function. The trend with redshift is
in the opposite direction to that in the \tcdm\ model except
at the highest redshift where the trend appears to reverse.  The initial
trend reflects the varying linking length
used to define the halos. For
\tcdm\ this choice was independent of redshift. Note that we use
natural logarithms in this plot.}  \figend

\figstart{figure=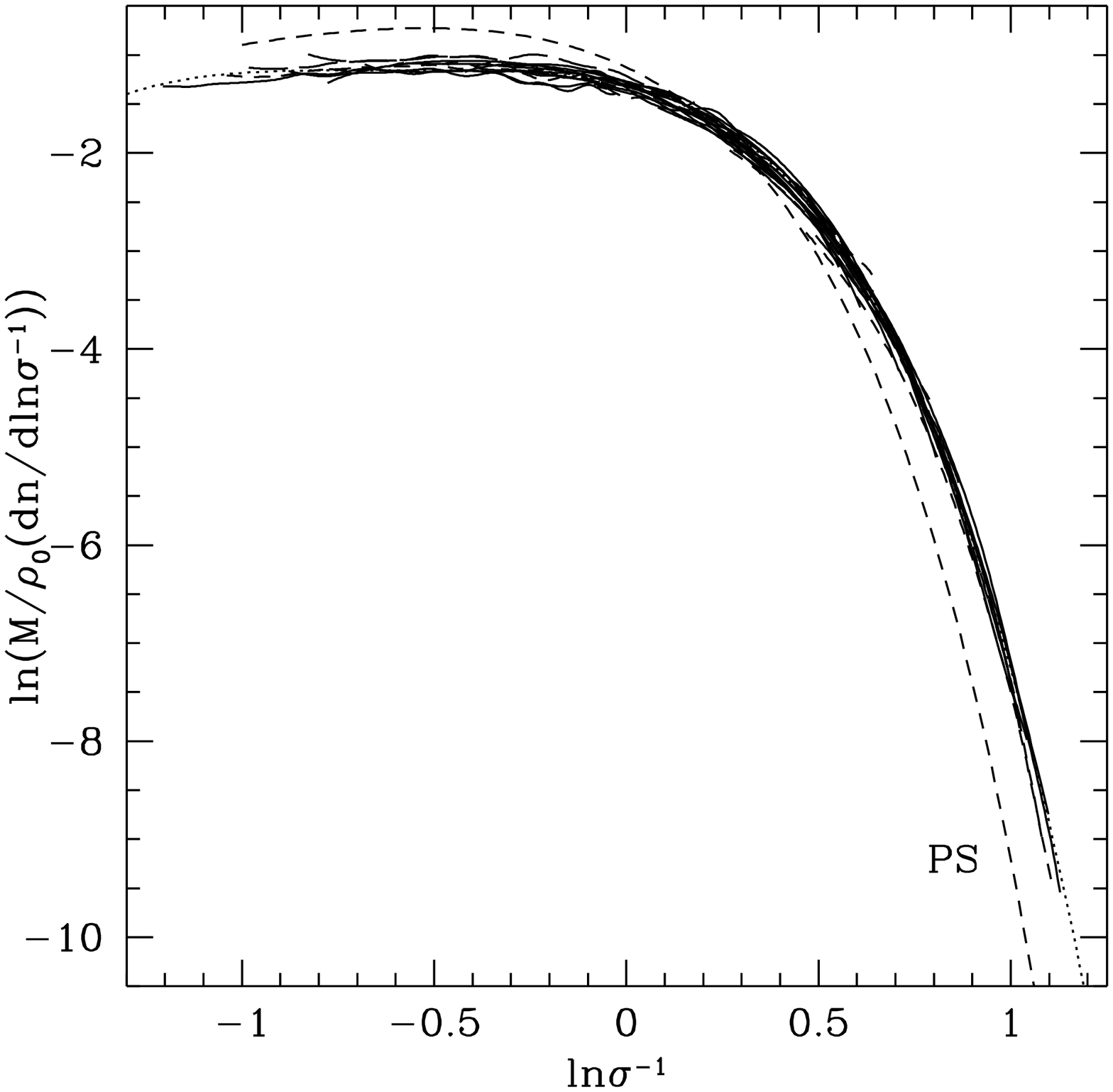,width=8.0cm}
\label{allfig}
\caption{The FOF(0.2) mass functions of all the simulation outputs
listed in Table~2.  Remarkably, when a single linking length is used
to identify halos at all times and in all cosmologies, the mass 
function appears to be invariant in the
$f-\ln\sigma^{-1}$ plane. A single formula (eqn.~\ref{tcdm_fit2}),
shown with a dotted line, fits all the mass functions with an accuracy
of better than about 20\% over the entire range. The dashed curve
show the Press-Schechter mass function for comparison.
}  \figend

Fig.~5 shows the mass functions determined from these two simulation
sets. We stress that the present comparison supersedes the preliminary
comparison presented by Governato et al. The
agreement between the various mass functions is good, with a 
variation of only 30\% over the range in $n_{\rm eff}$ from $-0.7$ to $-1.7$. 
On closer inspection, the curves from the \tcdm-hub simulation form a sequence
in which the mass function decreases with increasing redshift (or with
decreasing $n_{\rm eff}$) while the curves from Governato et al (1999)
also vary systematically, but in the opposite direction! The reason
for these differing weak trends is unclear. They could reflect the 
differences in power spectrum shape between the two simulations, or 
perhaps small systematic numerical errors in one or both of them.
(Note that since the Hubble volume simulation follows a much
larger volume than the simulation of Governato \etal, its mass function is
much better determined for large masses.)  However, the magnitude of these
trends (a total range below 30\%) is sufficiently small as to be of no
real interest.

Fig.~6 shows mass functions at various epochs in the \lcdm\ model.
Here the curves are affected not only by the variation of 
spectral shape with redshift (as for \tcdm) but also by
the variation in the linking length which defines the simulated halos.
For this plot we chose to follow the relation proposed by Eke \etal\ 
(1996). The $z=0$ mass function (the lowest solid curve in the figure)
is then significantly below the $z=0$ \tcdm\ mass function (the light 
dashed curve). Furthermore, in contrast to \tcdm, the
\lcdm\ mass function initially moves upwards in the $f-\ln\sigma^{-1}$
plane with increasing redshift. This can be attributed in large part
to the increasing linking length (see below). The upward trend reverses
between $z=0.96$ and $z=1.44$, perhaps reflecting the rapid
convergence of $\Omega$ and $\dot b/b$ to the \tcdm\ values.
As in the \tcdm\ case, the differences are all rather small. We conclude
that although with the halo definition used in this section, the
\tcdm\ and \lcdm\ mass functions are slightly different, there is no
evidence for a significant systematic variation in the high mass end
of the mass function with redshift, with power spectrum slope 
$n_{\rm eff}$, or with $\Omega$, once it is transformed to the
appropriate variables.

\subsection{A general fitting formula}\label{gen_mfn1}

The results of the last section suggest that it may be possible
to find a universal formula which provides a reasonably accurate
description of the halo mass function over a wide range of
redshifts and in a wide range of cosmologies. Here, we show that
a formula very similar to that suggested by S-T is indeed
successful if halos are defined at the same overdensity at all
times and in all cosmologies {\it independent of $\Omega$}. 
We concentrate on halos defined using FOF(0.2), but have found very
similar results using SO(180). Table~2 lists the simulation
outputs used in this section.  These include the data already analysed
from our own simulations and those of Governato \etal\ ,
together with additional data from the SCDM and OCDM (open with 
$\Omega_0=0.3$) simulations in Jenkins et al (1998).  Because of the
extended range of power spectral shapes involved, we `deconvolve' the
smoothed mass functions by multiplying by 
$\exp(-\alpha^2/59)$, where $\alpha$ is the local slope of $\log_{10}({\rm
d}n/{\rm d}\log_{10}M)$ (see Section~3).

Fig.~7 shows all the data plotted on the $f-\ln\sigma^{-1}$ plane.  As
before, all our curves are truncated at the mass corresponding to 20
particles (50 particles for the Governato \etal\ data) and where the
Poisson error first exceeds 10\%. These curves encompass a wide range
not only in $\ln\sigma^{-1}$ but also in effective power spectrum
slope, $n_{\rm eff}$. Cosmic density ranges over $0.3\le \Omega \le
1.0$.  Remarkably, all curves lie very close to a single locus in the
$f-\ln\sigma^{-1}$ plane.  The use of a constant linking length has
significantly reduced the amplitude of the redshift trend seen in the
\lcdm\ model in the previous section, and also places the \ocdm\
outputs on the same locus.

The numerical data in Fig.~7 are well fit by the following formula:
\begin{equation}
\label{tcdm_fit2}
 \phantom{xxxxxxx}f(M) =
 0.315\;\exp\big[-|\ln\sigma^{-1}+0.61|^{3.8}\big],
\end{equation}
valid over the range $-1.2\le\ln\sigma^{-1}\le1.05$.

\figstart{figure=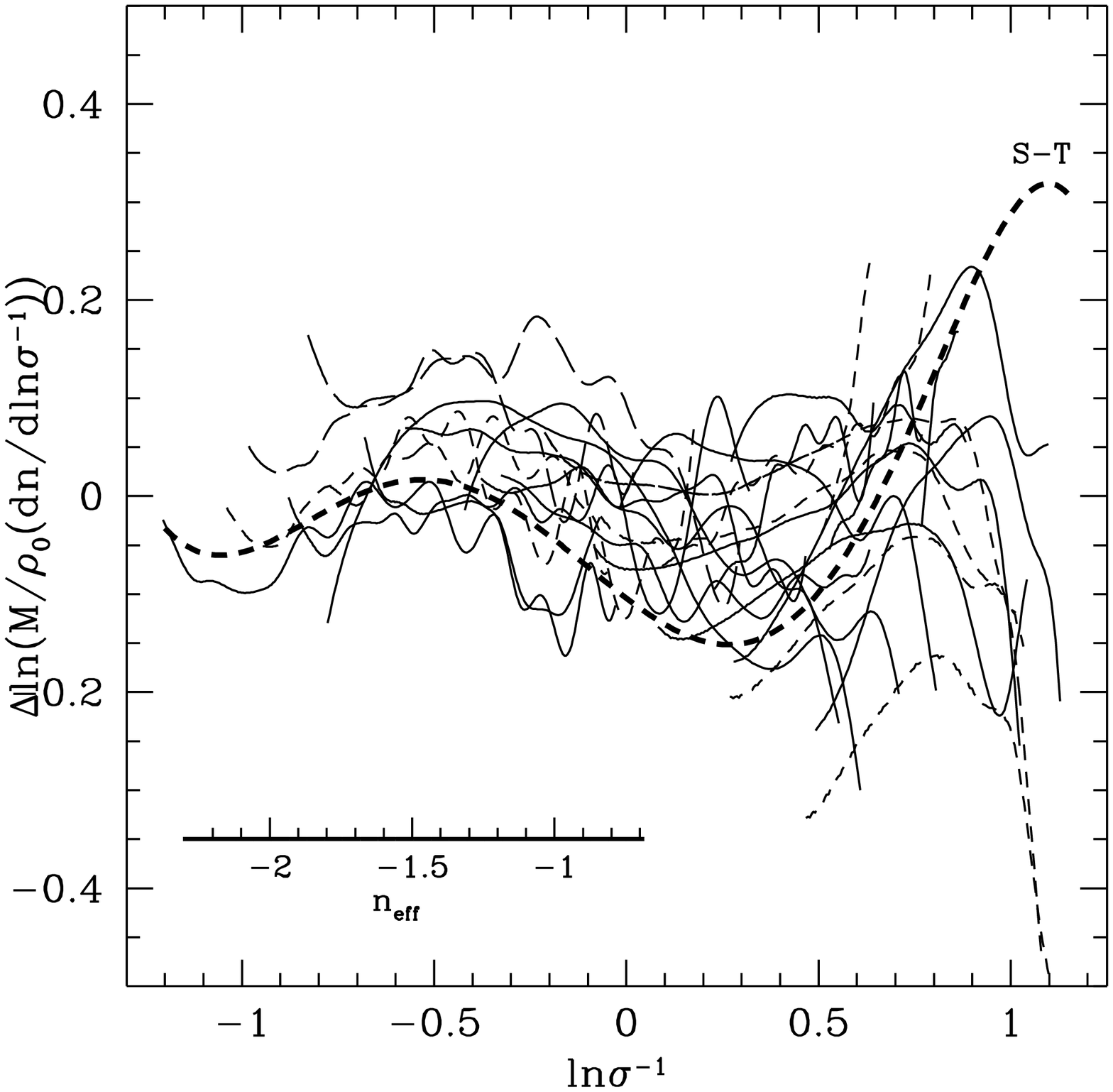,width=8.0cm}
\label{difffig}
\caption{The residual between the fitting formula, eqn.~\ref{tcdm_fit2},
and the FOF(0.2) mass functions for all the simulation outputs listed in
Table~2. The lines are colour codes according to the value of $n_{eff}$.
Solid lines correspond to simulations with $\Omega=1$, short
dashed lines to flat, low $\Omega_0$ models, and long dashed lines to open
models. The heavy dashed line shows the Sheth-Tormen formula (equation (7))
}  \figend

In Fig.~8 we plot the difference between the measured mass functions and
our fitting formula.  The fit is good to a fractional accuracy better than
20\% for $-1.2\le\ln\sigma^{-1}\le1$. This is a very significant
improvement over the Press-Schechter formula which would exceed the
vertical limits of the plot!  The curves for the open models with
$\Omega=0.3$ are slightly high in this plot but only by $\sim 10\%$.  The
spread between the different curves increases for large
$\ln\sigma^{-1}$. This may simply reflect the fact that the very steep high
mass end of the mass function is sensitive to numerical effects which
change the masses of clusters in a systematic way.

As shown in the figure, eqn.~9 is very close to the formula proposed by
Sheth \& Tormen (1999); there is a small difference in the high mass tail,
for $\ln\sigma^{-1}> 0.9$.  A non-linear least-squares fit of eqn.~7 to the
simulation data in Fig.~8 shows that the fit can be improved by adjusting
the parameters $A$, $p$ and $a$. If the normalisation constraint,
eqn.~6, is ignored, all three parameters can be allowed to vary freely. In
this case, the best fit is obtained for $A=0.353$, $p=0.175$ and $a=0.73$
(and 0.84 of the mass is in halos). If the normalisation constraint is
enforced, then only two parameters can vary; in this case the fit is not
as good as that provided by eqn.~9.

Fig.~9 shows the area of the $\ln\sigma^{-1}-n_{\rm eff}$ parameter 
space which is occupied by the data in Fig.~8.  The high mass end has good
coverage in $n_{\rm eff}$ with values up to -2.3. In practice this
means that for currently popular cosmologies, the high mass tail of
the halo mass function is well determined at all redshifts where
galaxies have so far been observed. The \tcdm-gif simulation
at $z=4.04$ has $n_{\rm eff}=-2.26$ and agrees well with 
\tcdm-hub which determine the high mass end of the mass function at more
recent epochs.  We have checked that the \tcdm-gif $z=5$ output, which has
$n_{\rm eff}=-2.35$, is also consistent with our fitting function, although
its Poisson errors are slightly too large to satisfy our 10\% criterion for
inclusion in Figs.~7--9.  For low $\Omega$ our fitting formulae should work
to even higher redshift. Since fluctuations grow more slowly for low
$\Omega$, and the value of $\sigma_8$ required to match current cluster
abundances is higher, low density cosmologies predict substantially less
negative values for $n_{\rm eff}$ at each redshift.

\begin{table*}
\centering
\begin{minipage}{140mm}
\caption{Parameters of N-body simulations used in Figs.~7,~8 and ~9. 
Columns~2 -- 5 give the cosmological parameters, and the normalisation of
the power spectrum at the present epoch. Columns~6 and~7 give the
number of particles and the size of the simulation cubes. Column~8 gives
the redshift of the output at which FOF groups were found, and
Column~9 gives the effective power spectrum slope at $\sigma=0.5$ defined
by equation~\ref{defnstar}.}
\begin{tabular}{@{}lrrlrrrrr@{}}
Simulation  &  $\Omega_0$ & $\Lambda_0$ & $\Gamma$ & $\sigma_8$ & 
$N_{\rm part}\phantom{aaa}$ 
     & $L_{\rm box}/h^{-1}{\rm Mpc}$ & $z$ & $n_{\rm eff}$ \\
\tcdm-gif &   1.0  & 0.0 & 0.21 & 0.60 & $16\,777\,216$ & $84.5\phantom{bbbb}$      &0.00  &  -1.39 \\
\tcdm-gif &   1.0  & 0.0 & 0.21 & 0.60 & $16\,777\,216$ & $84.5\phantom{bbbb}$      &1.94  &  -1.96 \\
\tcdm-gif &   1.0  & 0.0 & 0.21 & 0.60 & $16\,777\,216$ & $84.5\phantom{bbbb}$      &2.97  &  -2.13 \\
\tcdm-gif &   1.0  & 0.0 & 0.21 & 0.60 & $16\,777\,216$ & $84.5\phantom{bbbb}$      &4.04  &  -2.26 \\
\tcdm-int &   1.0  & 0.0 & 0.21 & 0.60 & $16\,777\,216$ & $160.3\phantom{bbbb}$     &0.00  &  -1.39 \\
\tcdm-virgo &   1.0  & 0.0 & 0.21 & 0.60 & $16\,777\,216$ & $239.5\phantom{bbbb}$   &0.00  &  -1.39 \\
\tcdm-512 &   1.0  & 0.0 & 0.21 & 0.51 & $134\,217\,728$ & $320.7\phantom{bbbb}$    &0.00  &  -1.48  \\
\tcdm-hub &   1.0  & 0.0 & 0.21 & 0.60 & $1\;000\,000\,000$ & $2000.0\phantom{bbbb}$&0.00  &  -1.39   \\
\tcdm-hub &   1.0  & 0.0 & 0.21 & 0.60 & $1\;000\,000\,000$ & $2000.0\phantom{bbbb}$&0.18  &  -1.48 \\
\tcdm-hub &   1.0  & 0.0 & 0.21 & 0.60 & $1\;000\,000\,000$ & $2000.0\phantom{bbbb}$&0.44  &  -1.58 \\
\tcdm-hub &   1.0  & 0.0 & 0.21 & 0.60 & $1\;000\,000\,000$ & $2000.0\phantom{bbbb}$&0.78  &  -1.70 \\
          &        &     &      &      &                    &                              &        \\
\scdm-gif &   1.0  & 0.0 & 0.50 & 0.60 & $16\,777\,216$ & $84.5\phantom{bbbb}$      &0.00  &  -0.99 \\
\scdm-virgo & 1.0  & 0.0 & 0.50 & 0.60 & $16\,777\,216$ & $239.5\phantom{bbbb}$     &0.00  &  -1.08 \\
          &        &     &      &      &                    &                              &        \\
\lcdm-gif & 0.3  & 0.7 & 0.21& 0.90 & $16\,777\,216$ & $141.3\phantom{bbbb}$        &0.00  &  -1.17 \\
\lcdm-gif & 0.3  & 0.7 & 0.21& 0.90 & $16\,777\,216$ & $141.3\phantom{bbbb}$        &0.52  &  -1.32 \\
\lcdm-gif & 0.3  & 0.7 & 0.21& 0.90 & $16\,777\,216$ & $141.3\phantom{bbbb}$        &2.97  &  -1.72 \\
\lcdm-gif & 0.3  & 0.7 & 0.21& 0.90 & $16\,777\,216$ & $141.3\phantom{bbbb}$        &5.03  &  -2.00 \\
\lcdm-512 &   1.0  & 0.0 & 0.21 & 0.90 & $134\,217\,728$ & $479.0\phantom{bbbb}$    &0.00  &  -1.25  \\
\lcdm-512 &   1.0  & 0.0 & 0.21 & 0.90 & $134\,217\,728$ & $479.0\phantom{bbbb}$    &5.00  &  -2.08  \\
\lcdm-hub & 0.3  & 0.7 & 0.21& 0.90 & $1\,000\,000\,000$ & $3000.0\phantom{bbbb}$   &0.00  &  -1.25  \\
\lcdm-hub & 0.3  & 0.7 & 0.21& 0.90 & $1\,000\,000\,000$ & $3000.0\phantom{bbbb}$   &0.27  &  -1.32  \\
\lcdm-hub & 0.3  & 0.7 & 0.21& 0.90 & $1\,000\,000\,000$ & $3000.0\phantom{bbbb}$   &0.96  &  -1.51  \\
\lcdm-hub & 0.3  & 0.7 & 0.21& 0.90 & $1\,000\,000\,000$ & $3000.0\phantom{bbbb}$   &1.45  &  -1.62  \\
          &        &     &      &      &                    &                              &        \\
\ocdm-gif &   0.3  & 0.0 & 0.21 & 0.85 & $16\,777\,216$ & $141.3\phantom{bbbb}$      &0.00  &  -1.20 \\
\ocdm-virgo & 0.3  & 0.0 & 0.21 & 0.85 & $16\,777\,216$ & $239.5\phantom{bbbb}$     &0.00  &  -1.20 \\
          &        &     &      &      &                    &                              &        \\
\scdm-gov &   1.0  & 0.0 & 0.5 & 1.0 & $46\,656\,000$ & $500.0\phantom{bbbb}$      &0.00  &  -0.71 \\
\scdm-gov &   1.0  & 0.0 & 0.5 & 1.0 & $46\,656\,000$ & $500.0\phantom{bbbb}$      &0.43  &  -0.90 \\
\scdm-gov &   1.0  & 0.0 & 0.5 & 1.0 & $46\,656\,000$ & $500.0\phantom{bbbb}$      &1.13  &  -1.12 \\
\scdm-gov &   1.0  & 0.0 & 0.5 & 1.0 & $46\,656\,000$ & $500.0\phantom{bbbb}$      &1.85  &  -1.28 \\
\end{tabular}
\end{minipage}
\end{table*}

\figstart{figure=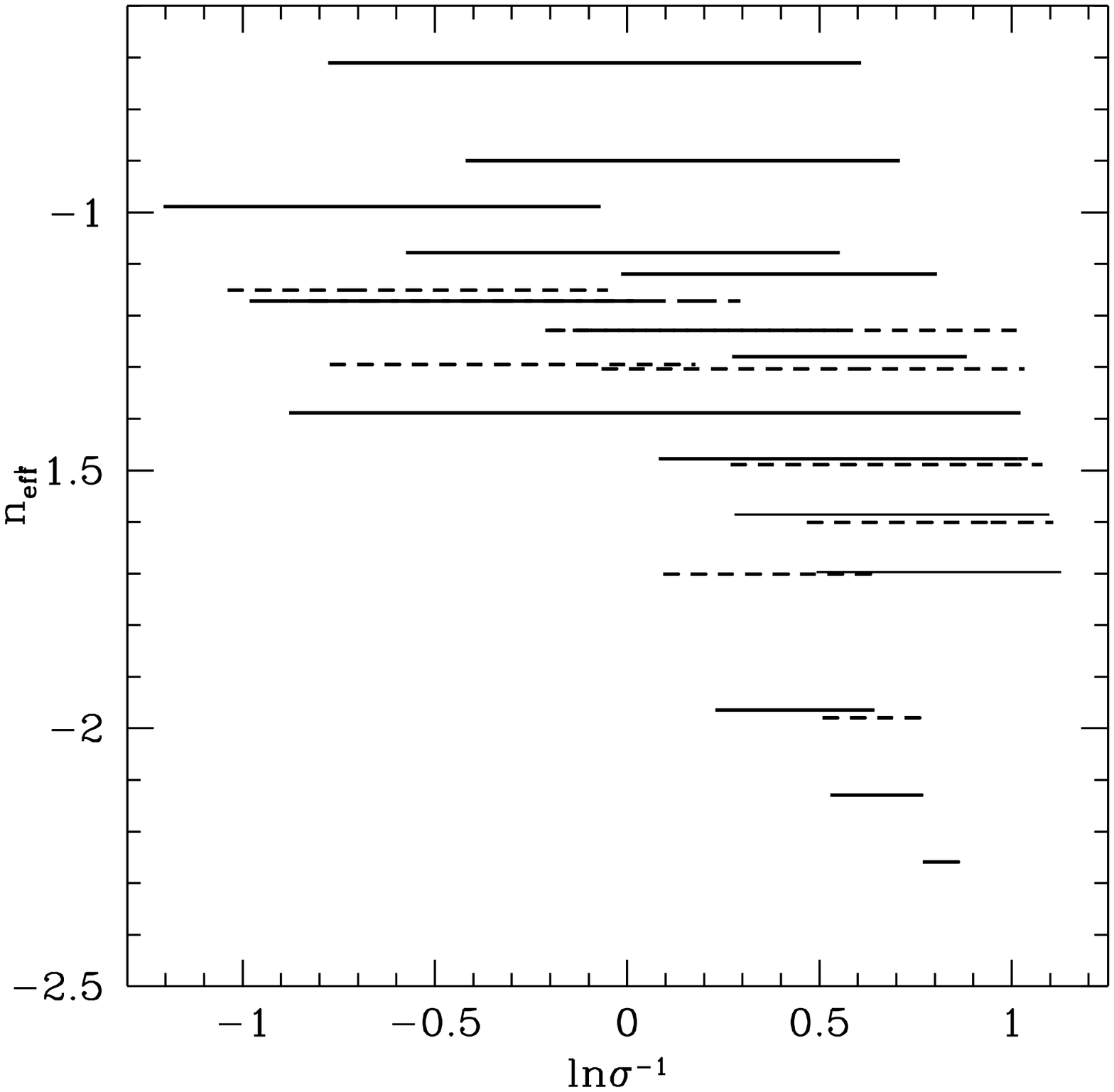,width=8.0cm}
\label{nstarf}
\caption{The parameter range covered by the simulation outputs listed in 
Table~2. The high mass end of the mass function is well determined for a
range of values of the parameter $n_{\rm eff}$, while the low mass end is
only determined well for high values of $n_{\rm eff}$. Dashed lines
indicate models with $\Omega_0<1$.}
\figend

\section{Conclusions}\label{cc}

We have derived halo mass functions at $z=0$ from simulations 
of the \tcdm\ and \lcdm\ cosmologies over more than
four orders of magnitude in mass, $\sim 3\times
10^{11}$ to $\sim5\times10^{15 }\hmsun$. In particular, our two Hubble
volume simulations provide the best available predictions for the
abundance of the most massive clusters.  We have checked the
sensitivity of our mass functions to choice of group-finder, to
limiting overdensity, and to numerical parameters such as softening,
particle mass and starting redshift (see appendix A). Most dependences
are weak. In particular, with a friends-of-friends group finder,
the mass function is robustly determined with systematic uncertainties
at or below the 10\% level for groups containing 20 particles or
more. Somewhat higher particle numbers are needed for reliable results
with a spherical overdensity group-finder. 

The mass functions we find for these two cosmologies, as well as for 
additional simulations of the SCDM and OCDM cosmologies, display the
kind of universality predicted by the Press-Schechter model. When
expressed in suitable variables, the mass function is independent of
redshift, power spectrum shape, $\Omega$ and $\Lambda$. This 
universality only obtains when we define halos in our simulations 
at fixed overdensity {\it independent} of $\Omega$. When we use
the spherical collapse model to define the appropriate 
overdensity, as suggested by Lacey \& Cole (1994) and Eke \etal\ (1996),
we find mass functions for low density cosmologies which vary weakly 
but systematically with redshift. As has been noted before, the 
Press-Schechter model overestimates the abundance of $M_*$ halos 
and underestimates the abundance of massive halos in all cosmologies.
On the other hand, the fitting function proposed by Sheth \& Tormen 
(1999) is a very good fit to the universal mass function we find, and 
is close to the best fit we give as equation (9). As shown by Sheth,
Mo \& Tormen (1999) this shape is a plausible consequence of extending
the excursion set derivation of the P-S model to include ellipsoidal
collapse.

Our ``universal'' mass function has considerable generality since 
the simulations cover a wide
range of parameter space: $\Omega$ in the range $0.3 - 1$, effective spectral
power index in the range $-1$ to $-2.5$, inverse fluctuation amplitude in the
range $-1.2\le\ln\sigma^{-1}\le1.05$. For standard cosmologies
this corresponds approximately to the mass range $10^{11}$ to 
$10^{16}\hmsun$ at $z=0$, and to the high mass tail of the mass
function out to redshift 5 or more. More work is needed to check
the abundances predicted for low mass halos, in particular to see
whether all mass is predicted to be part of some halo, or
whether some fraction makes up a truly diffuse medium.

Data for many of the simulations analysed in this paper,
as well as cluster and galaxy catalogs created as part of 
other projects, are available from http://www.mpa-garching.mpg/Virgo

Software to convert eqn.~(\ref{tcdm_fit2}) into a mass function, for a
given power spectrum is available on request from ARJ
(A.R.Jenkins@durham.ac.uk).

\section*{Acknowledgements}

The simulations discussed here were carried out as part of the Virgo
consortium programme, on the Cray-T3D/Es at the Edinburgh Parallel
Computing Centre and the Rechenzentrum, Garching.  We thank Fabio Governato
for supplying the halo catalogues for the simulation of Governato et al
(1999).  This work was supported by the EC network for ``Galaxy formation
and evolution'' and NATO CRG 970081 and by PPARC. CSF acknowledges a PPARC
Senior Research Fellowship and a Leverhume Research Fellowship. SMC
acknowledges a PPARC advanced fellowship.
 
We thank the referee for his constructive comments on the manuscript.

\appendix
\section{A Resolution Study}
\figstart{figure=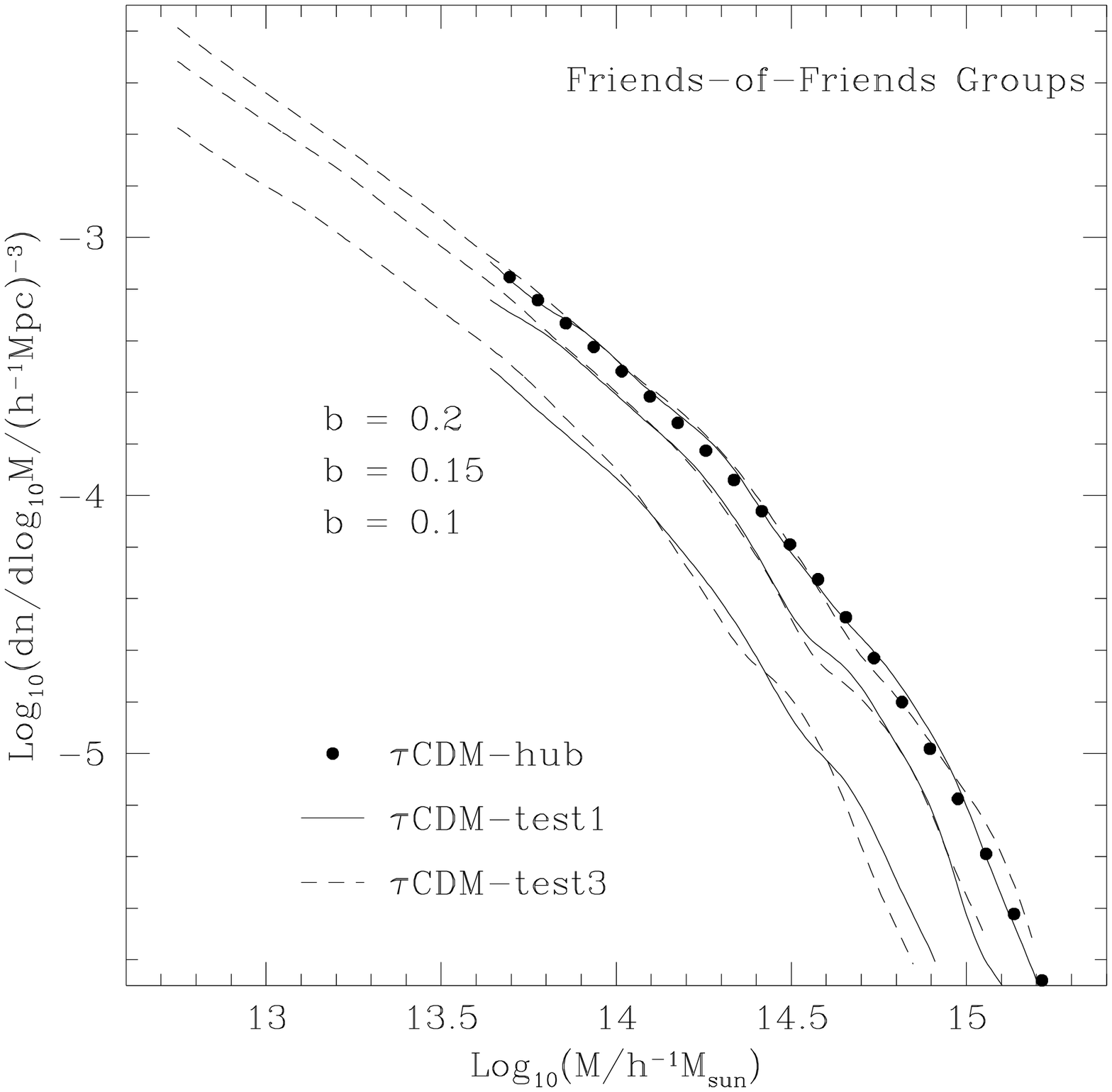,width=8.0cm}
\label{figure-a1}
\caption{A resolution study of the effect of varying the particle mass on
the mass function of FOF halos for three values of the linking parameter,
$b$. The dashed lines show the mass functions obtained from \tcdm-test3,
and the solid lines from \tcdm-test1. These simulations have the same
phases but differ by a factor of 8 in particle mass.  The mass function is
plotted for halos with 20 particles and above. The filled circles show the
corresponding mass function in \tcdm-hub which has the same particle mass
as \tcdm-test1. Defined in this way, the mass function is remarkably
robust. } \figend

\figstart{figure=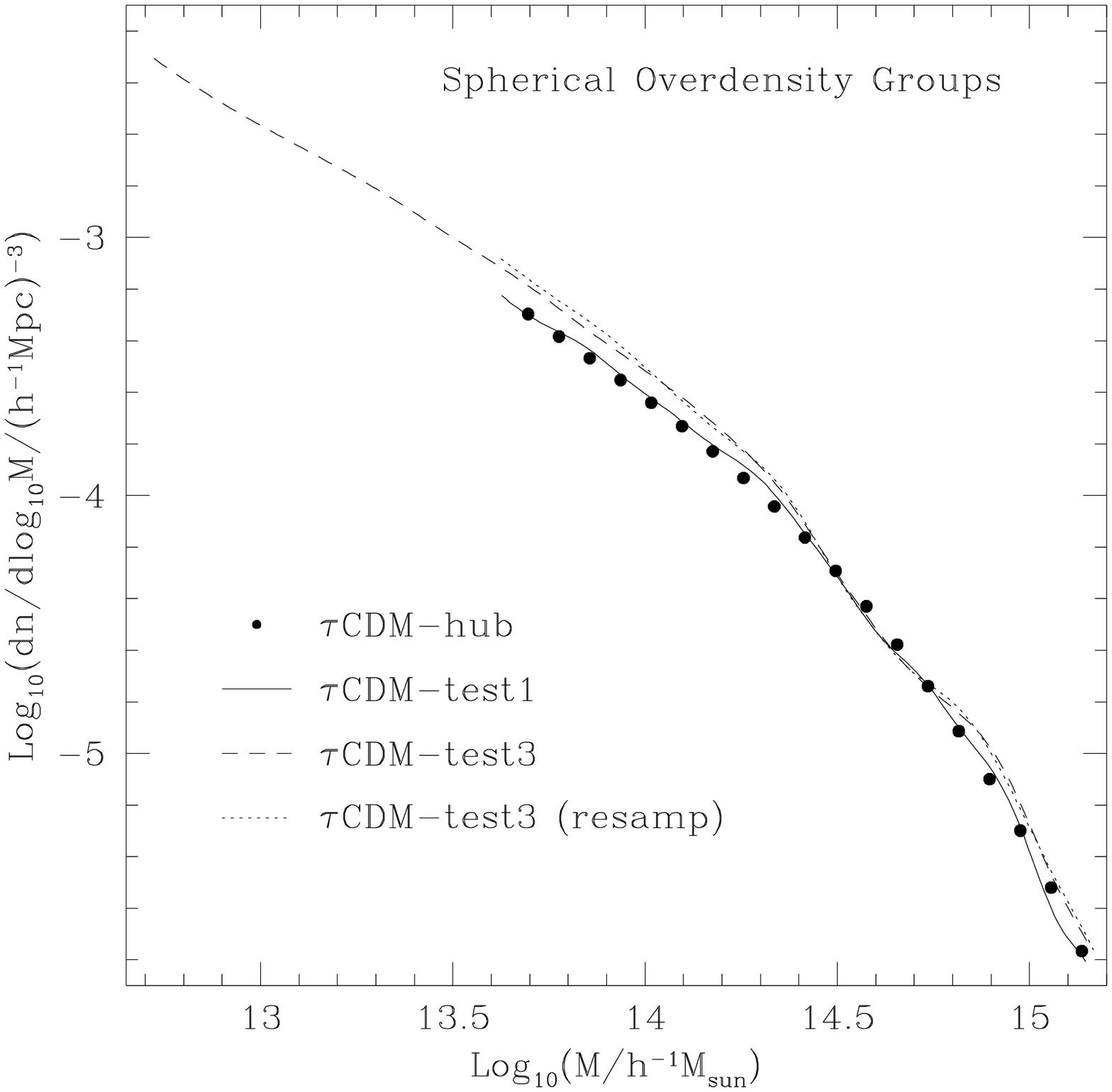,width=8.0cm}
\label{figure-a2}
\caption{A resolution study of the effect of varying the particle mass 
on the mass function of SO halos. The dashed line shows the mass
function obtained from \tcdm-test3, and the solid line from
\tcdm-test1. These simulations have the same phases but differ by a
factor of 8 in particle mass.  The mass function is plotted for halos
with 20 particles and above. The filled circles show the corresponding
mass function in \tcdm-hub which has the same particle mass as
\tcdm-test1. The dotted line shows the SO mass function when 1 in 8
particles are randomly sampled from \tcdm-test3. This test shows that
the SO mass function is sensitive to the particle mass, unlike the FOF
mass function in Fig.~A1. } \figend

To investigate the effects of mass resolution, gravitational softening
and starting redshift, we carried out the three test simulations
detailed at the bottom of Table~1. These simulations are all of an
identically sized region, $200\hmpc$ on a side, and are set up so that
corresponding waves have identical phases and amplitudes in all three
cases. Two of the simulations, \tcdm-test1 and \tcdm-test2, have the
same particle mass as the Hubble volume simulation. The former has
essentially identical numerical parameters to the \tcdm-hub
simulation, while the latter differs in having a smaller gravitational
softening length.  The \tcdm-test3 simulation has 8 times as many
particles as \tcdm-test1, but is otherwise identical. As a check of
the effect of the starting redshift we repeated
\tcdm-test1 a second time but starting at redshift 14 rather than 29 (as
for the other tests and \tcdm-hub).

Fig.~A1 shows the mass functions for halos found with the FOF group-finder
using three different linking lengths ($b=0.1,0.15$ and 0.2), for
\tcdm-test1 and \tcdm-test3. Consistent with the results of Section~3, the
FOF mass function is only very weakly dependent on the particle mass.
Changing the softening makes an even smaller difference and we do not plot
the curve for \tcdm-test2; the average RMS difference between the mass
functions with 30\hkpc\ and 100\hkpc\ gravitational softening is just
0.0135 dex. If the softening were increased significantly beyond 100\hkpc,
then we would expect the mass function to decrease as halos become more and
more diffuse.  The mass function for \tcdm-hub\ is shown also and agrees
well with \tcdm-test1, as it should. We conclude that the mass function of
FOF halos is remarkably robust to the numerical parameters in the
simulations.

When halos are identified using the SO group-finder, the situation is
slightly different.  As Fig.~A2 shows, the SO mass function from
\tcdm-test1\ agrees well with that in \tcdm-hub over the corresponding
range of masses.  Changing the softening to 30\hkpc\ in \tcdm-test2\
does not change the mass function much.  However, changing the mass
resolution by a factor of 8, as in \tcdm-test3, leads to a significant
change in the mass function at the low mass end, which is now much
closer to that determined from \tcdm-virgo, a simulation with similar
mass resolution. Resampling the high-mass resolution simulation,
\tcdm-test3, at random, at a rate of 1-in-8 does not have any
noticeable effect on the mass function. This suggests that the
resolution-dependence of the mass function close to the resolution limit
is not due simply to
details of the cluster-finding algorithm (such as the position of the
halo centre). Rather, it suggests that the difference may reflect
genuine differences in the structural properties of marginally
resolved halos in simulations with different particle numbers.

The mass function is not very sensitive to the starting redshift unless
this is so low that the neighbouring Zel'dovich displacements are so large
as to interfere with the formation of the first non-linear objects. In
practice, all the simulations compiled here pass this test easily. As a
additional check, we repeated \tcdm-test1, starting at $z=14$. This made
very little difference to the mass function overall, causing an RMS average
difference over the measured mass function of only 0.02 dex.

\begin{figure*}
\centerline{\psfig{figure=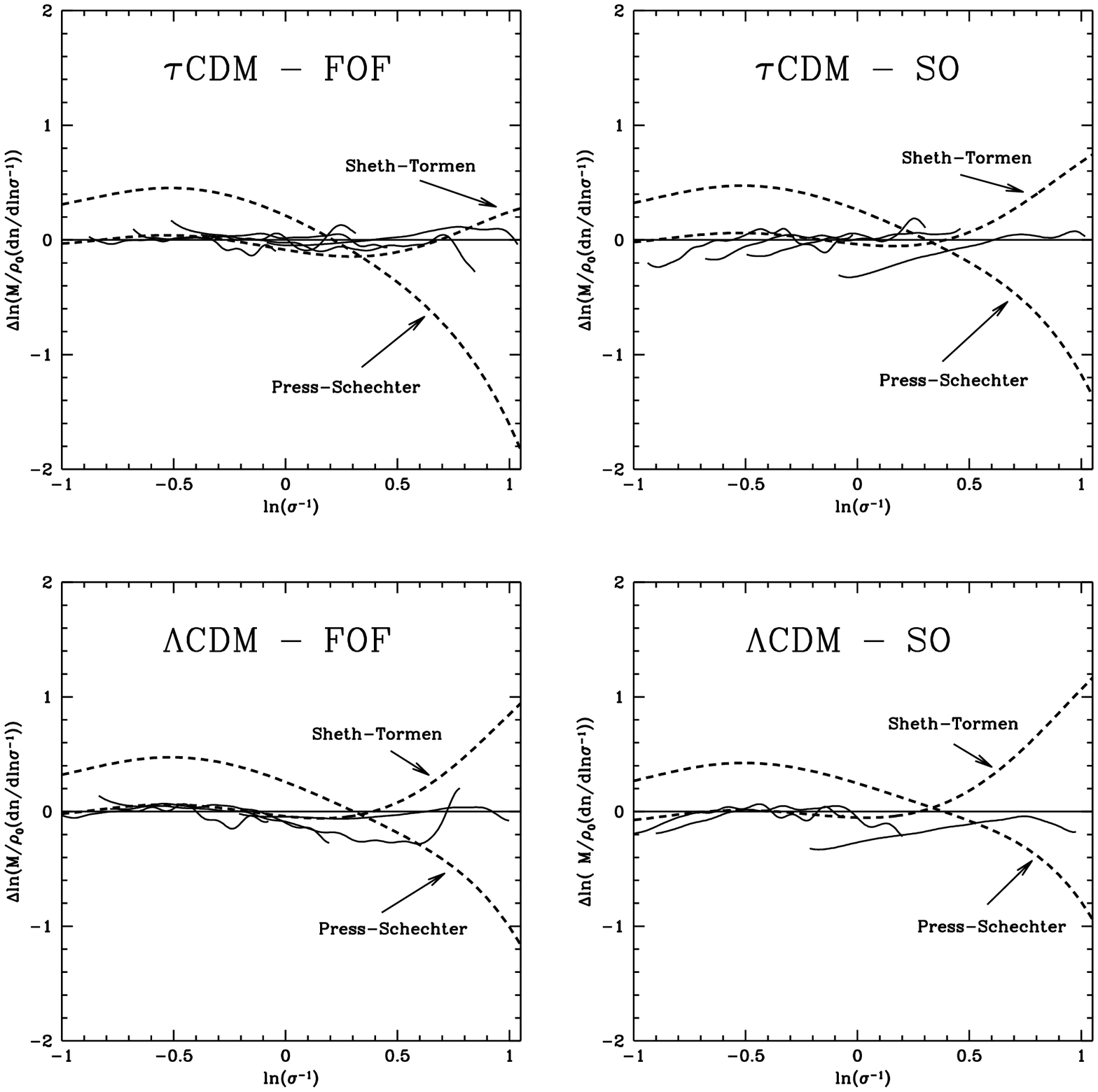,width=400pt,height=400pt}}
\caption{The residuals between the fitting formulae for the mass
function given in appendix B and (i) the simulations (solid lines)
and (ii) the Press-Schechter and Sheth-Tormen models (dashed
lines.)  The simulation curves are plotted up to the point where the
fractional Poisson error reaches 10\%.  The simulations themselves
can be seen to be mutually consistent at about the 10\% level although
the differences are, for the most part, larger than the Poisson 
errors.  The P-S curve shows an excess at low masses and a deficit at
high masses. The Sheth-Tormen mass function fits the low mass end
well, but overestimates the number of high mass clusters. Note
that we use natural logarithms in this plot.}
\end{figure*}

In summary, our tests indicate that we can derive an accurate mass
function for FOF groups from the Virgo simulations (including the
Hubble volume). The mass functions agree well in the overlap regions
of simulations of different mass resolution even when halos with only
20 particles per group are included. Perhaps surprisingly, the mass
functions of SO groups do not match up nearly
as well at such low particle numbers. A minimum of
$\sim 100$ particles is required to provide reasonable agreement in the
overlap regions in this case.

\section{Fitting formulae for the FOF and SO mass functions}

Here, we give fitting formulae for the (unsmoothed) mass functions 
of FOF and SO halos, using the standard values of the
group-finding parameters given in Section~2. These fits are plotted in 
Figs.~3 and~4 of Section~4. 

1. {\it The friends-of-friends group finder}: 

\noindent\tcdm/FOF(0.2):
\begin{equation}
\label{tcdm_fit}
 \phantom{xxxxxxx}f(M) = 
 0.307\;\exp\big[-|\ln\sigma^{-1}+0.61|^{3.82}\big],
\end{equation}
in the range $-0.9\le\ln\sigma^{-1}\le1.0$. 

\noindent\lcdm/FOF(0.164):
\begin{equation}\label{lcdm_fit}
 \phantom{xxxxxxx} f(M) = 
     0.301\;\exp\big[-|\ln\sigma^{-1}+0.64|^{3.88}\big],
\end{equation}
in the range $-0.96\le\ln\sigma^{-1}\le1.0$.  The difference between
these fitting functions and the actual mass functions is typically
less than 10\%.

2. {\it The spherical overdensity group finder:} 

\noindent\tcdm/SO(180):
\begin{equation}\label{tcdm_so_fit}
 \phantom{xxxxxxx}f(M) = 
     0.301\;\exp\big[-|\ln\sigma^{-1}+0.64|^{3.82}\big],
\end{equation}
for the range $-0.5\le\ln\sigma^{-1}\le1.0$.

\noindent\lcdm/SO(324): 
\begin{equation}\label{lcdm_so_fit}
 \phantom{xxxxxxx}f(M) = 
     0.316\;\exp\big[-|\ln\sigma^{-1}+0.67|^{3.82}\big],
\end{equation}
for the range $-0.7\le\ln\sigma^{-1}\le1.0$.  

As discussed earlier, the differences between the SO mass functions
are larger than for FOF halos. Differences between
the fit and the mass functions within the quoted mass range can be as large
as 20\%.

Fig.~B1 shows the residuals between the four fitting formulae quoted above
and the mass functions in the simulations. Also shown are the differences
between our fitting formulae and both the P-S and S-T models.  The simulation
curves are truncated at the high mass end at the point where the fractional
RMS Poisson error reaches 10\%. The simulation mass functions may be seen
to be consistent with one another at a level of about 10\%.  As seen in
previous plots, the P-S curve is too high at low masses (low
$\ln\sigma^{-1}$) and too low at high masses. For the fit proposed by
Sheth \& Tormen (1999, eqn.~10) good agreement is to be expected at the
low mass end since a subset of the simulation data used here (the -gif
simulations) was used by Sheth \& Tormen (1999). Although their
mass function is normalised (so that all the mass is attached to halos), it
does match our fits for \tcdm-FOF rather well over their entire range. For
the other models, the Sheth-Tormen fit overestimates the number of very
massive clusters.  Because of the normalisation constraint, the
Sheth-Tormen mass function exceeds the P-S predictions for extremely low
mass halos but this occurs at a point well beyond what can be tested easily
in N-body simulations.  Whether such low-mass behaviour is correct 
remains to be determined.

\label{lastpage}
\end{document}